\documentclass{article}

\usepackage{amsmath}
\usepackage{amssymb}
\usepackage{enumitem}
\usepackage{young}
\usepackage{ytableau}
\usepackage{graphicx}
\usepackage{cool}
\usepackage{hhline}
\usepackage{tabularx}
\usepackage{enumitem}   
\usepackage[hidelinks=false]{hyperref}
\usepackage{color}
\usepackage{kantlipsum,setspace}
\usepackage{subcaption}
\usepackage[margin=1.5in]{geometry}
\usepackage{authblk}
\makeindex

\newcommand{\NAH}{Hermanns hierarchy}
\newcommand{\su}[2]{\mathrm{su}\!\left(#1\right)_{#2}}
\newcommand{\nsu}[2]{\left(#1,#2\right)}
\newcommand{\pf}[3]{\mathrm{su}\!\left(#1\right)_{#2}/\mathrm{u}\!\left(1\right)^{#3}}
\newcommand{\bg}{\mathcal{O}_{\mathrm{bg}}}
\newcommand{\bb}[1]{\langle #1 X \rangle }

\begin{document}
\title{Braiding properties of paired spin-singlet and non-abelian hierarchy states}
\author{Yoran Tournois and Eddy Ardonne}
\affil{\it{Department of Physics, Stockholm University, SE-106 91 Stockholm, Sweden}}
\date{}
\maketitle
\subsubsection*{Abstract}
We study explicit model wave functions describing the fundamental quasiholes in a class of non-abelian fractional quantum Hall states. This class is a family of paired spin-singlet states with $n\geq1$ internal degrees of freedom. We determine the braid statistics of the quasiholes by determining the monodromy of the explicit quasihole wave functions, that is how they transform under exchanges of quasihole coordinates. The statistics is shown to be the same as that of the quasiholes in the Read-Rezayi states, up to a phase. We also discuss the application of this result to a class of non-abelian hierarchy wave functions.

\section{Introduction}
\label{sec:introduction}
The discovery of the fractional quantum Hall effect \cite{Tsui1982} has led to the prediction of fractionally charged quasiparticle excitations \cite{Laughlin1983}, quasiholes and quasielectrons, obeying fractional statistics \cite{Leinaas1977,Wilczek1982}. For most quantum Hall states the quasiparticle statistics is expected to be abelian, i.e. the many-quasiparticle wave function picks up a fractional phase under the exchange of the quasiparticle coordinates. However, certain states are thought to host non-abelian excitations \cite{Moore1991}, in which case the many-quasiparticle wave function has multiple components which transform according to a unitary braid matrix $U_{ij}$ when quasiparticles at positions $w_i$ and $w_j$ are exchanged. 

One way in which the theoretical understanding of the fractional quantum Hall effect has progressed is by proposing trial wave functions for ground states and excited states, with the goal of capturing topological properties such as the fractional charges and braiding statistics of the quasiparticle excitations. Examples of abelian trial wave functions are the Laughlin wave function \cite{Laughlin1983}, the hierarchy wave functions \cite{Haldane1983,Halperin1984} and the composite fermion (CF) wave functions \cite{Jain1989}; examples of non-abelian wave functions include the Moore-Read \cite{Moore1991} and, more generally, the Read-Rezayi \cite{Read1999} series.
The Moore-Read state, or rather its particle-hole conjugate, the `anti-pfaffian' \cite{levin-anti-pfaffian07,lee-anti-pfaffian07}, are leading candidates for describing the plateau at $\nu=\frac{5}{2}$ based on numerical studies, see for instance \cite{rezayi-52-numerics11}.

A powerful tool in proposing and analyzing such model wave functions has been conformal field theory (CFT). Various trial wave functions can be expressed as conformal blocks and it was conjectured that this description makes the topological properties, in particular the braiding properties, of the wave function manifest \cite{Moore1991}. The braiding statistics of quasiholes is represented by the Berry holonomy which has contributions from Berry phase accumulated during the exchange as well as the explicit transformation -- the monodromy --  of the wave function \cite{Blok1992}. For the Laughlin \cite{Kjonsberg1997,Arovas1984} as well as the Moore-Read case  \cite{Bonderson2011} (among other ``Ising type'' states, see also  \cite{Gurarie1997,Read2009b}) it was shown that the CFT description is one in which the statistics is given by the monodromy, with a trivial Berry phase. This was verified numerically in the Laughlin case \cite{Kjonsberg1999}, and in the Moore-Read and $\mathbb{Z}_3$ Read-Rezayi \cite{Wu2014} cases using the matrix product state formulation of~\cite{Zaletel2012}. In those cases, therefore, the braid statistics of quasiholes can be inferred from the manifest transformation of the quasihole wave function. 

In this paper, we study the braiding properties of quasiholes in a one-parameter family of non-abelian model wave functions denoted $\Psi_{(n+1,2)}$, with $n\geq 1$. Referred to as paired spin-singlet states, this family is a generalization of the spin polarized Moore-Read wave function ($n=1$) and the non-abelian spin-singlet (NASS) \cite{Ardonne1999} wave function ($n=2$), to particles carrying $n$ quantum numbers determining the charge and (pseudo-) spin. Such model wave functions have been considered in the context of rotating spin-1 bosons for $n=3$ \cite{Reijnders2002,Reijnders2004}, graphene \cite{vanvoorden2016}, as well as fractional Chern insulators \cite{Sterdyniak2013,Zhao2013} with Chern number $C>1$. Related wave functions were studied in \cite{wen91,Blok1992,wen99} using a parton construction. Recently, progress was made on the Landau-Ginzburg theories describing these states \cite{goldman2019}.

According to the `Moore-Read conjecture' \cite{Moore1991} (see \cite{Hansson2017} for a review) the CFT representation of the paired spin-singlet states should make the braiding properties manifest in the monodromy. By finding explicit quasihole wave functions, the braid matrices for the Moore-Read wave functions were found in \cite{Nayak1996}, and those for the Read-Rezayi and NASS cases were determined in \cite{Ardonne2007}. We study the manifest transformation properties of the paired spin-singlet states by obtaining explicit expressions for four-quasihole wave functions using conformal field theory techniques. This calculation relies on explicit four-point functions in certain Wess-Zumino-Witten (WZW) models which were obtained in Ref. \cite{Knizhnik1984}, as well as the properties of the closely related parafermion CFTs \cite{Gepner1987} which are presented in Appendix \ref{sec:appendix_pf}. We show that the braiding properties of the quasiholes for $\Psi_{(n+1,2)}$ are, up to a phase, the same as those of the quasiholes in the $\mathbb{Z}_{n+1}$ Read-Rezayi states \cite{Read1999}, which reflects the rank-level duality between their CFT descriptions. 

The paired spin-singlet states are also closely related to a set of non-abelian hierarchy wave functions proposed in \cite{Hermanns2010} based on a picture of successive condensation of non-abelian quasiparticles. This set of trial wave functions, which we refer to as Hermanns hierarchy wave functions, can be thought of as bilayer composite fermion wave functions where one performs a symmetrization (or antisymmetrization) over the layer index. These have been studied numerically in Ref. \cite{Sreejith2011}, showing that they are promising candidates for the second Landau level. The simplest (non-trivial) \NAH{} state was shown  to be closely related to the non-abelian spin-singlet state \cite{Hermanns2010}; in Ref. \cite{Tournois2017} it was shown that the other \NAH{} states are similarly related to the paired spin-singlet states. Using this relation, we argue that the braiding properties of quasiholes in the \NAH{} states should be the same as those in the paired spin-singlet states.

The paper is organized as follows. In Section \ref{sec:fqhe_cft}, we briefly review the connection between trial wave functions in the fractional quantum Hall effect and conformal field theory. In Section \ref{sec:paired_spin_singlets} we discuss the paired spin-singlet states in detail and introduce `master formulas' that relate two representations of the paired spin-singlet states, which allows us to find explicit wave functions for four quasiholes. In Section \ref{sec:n=3_braiding}, we present the calculation of the braiding properties in the paired spin-singlet state $\Psi_{\left(4,2\right)}$ after which we present the calculation for a general paired spin-singlet state in Section \ref{sec:r_braiding}. Finally, in Section \ref{sec:application}, we comment on the relation between the paired spin-singlet states and the \NAH{} states. In the appendices, we provide details on the WZW CFTs, the associated parafermion CFTs and the consequences of rank-level duality for the braid matrices studied in this paper.

\section{Model wave functions and conformal field theory}
\label{sec:fqhe_cft}
We consider model wave functions for fractional quantum Hall states of the form
\begin{equation}
\begin{aligned}
\Psi^M \left(z_1,\ldots,z_N\right) &=  \Phi \left(z_1,\ldots,z_N\right) \prod_{i<j} \left(z_i-z_j\right)^M e^{-\frac{1}{4}\sum_i \left|z_i\right|^2},
\label{eq:model_wf}
\end{aligned}
\end{equation}
where $M\geq 0 $ and the magnetic length $l_B$ has been set to 1. In Eq. \eqref{eq:model_wf}, $\Phi$ is a \emph{symmetric} holomorphic function of the particle coordinates $z_j=x_j+iy_j$, obeying certain vanishing conditions. In this paper we only consider paired states, for which 
$\Phi\left(z_1,\dots,z_N\right)\big|_{z_i=z_j}\neq0$, but $\Phi\left(z_1,\ldots,z_N\right)\big|_{z_i=z_j=z_k}=0$ for any distinct $z_i,z_j,z_k$. The wave function $\Psi^{M=0}$ is bosonic and has the same pairing property, while the simplest fermionic wave function corresponds to $M=1$. The power $M$ of the Vandermonde determinant $\prod_{i<j} \left(z_i-z_j\right)^M$ is chosen maximally, that is $\Phi \left(z_1,\ldots,z_N\right)$ is the polynomial of lowest degree with the pairing property as described above. To simplify the discussion we set $M=0$ from here on, denoting $\Psi^{M=0}$ by $\Psi$. We will consider the general wave functions with $M>0$ at a later stage. We also suppress the Gaussian factors.

In the following we make extensive use of the connection between CFT and the fractional quantum Hall effect \cite{Moore1991,Frohlich2001}, by means of which trial wave functions are expressed as (chiral) conformal blocks in a certain CFT. In particular, the wave function is represented by a vacuum expectation value of (radially ordered) operators in the CFT which describe the constituent (quasi)particles. The trial wave function for the ground state reads
\begin{equation}
\Psi (z_1,\ldots,z_N) = \langle  \bg V(z_1) \cdots V(z_N)\rangle.
\label{eq:wf_cft}
\end{equation}
Here the operator $V$ represents an electron\footnote{Although the particles described by $V$ are bosons for $M=0$, we refer to them as electrons.} and the operator $\bg$ is a background charge operator which is needed to ensure a nonzero result: it can chosen in such a way that the Gaussian factors are reproduced \cite{Moore1991}. By a simple change of the operators $V$, the wave function $\Psi^M$ for general $M$ can also be represented in this way. Similarly, model wave functions for quasiholes can be obtained by including appropriate operators $H$ at  positions $w=w_x+iw_y$.

Not all CFTs give appropriate trial wave functions: there are certain conditions to be satisfied \cite{Frohlich2001,Read2009}, most notably the existence of appropriate operators to represent the electrons and quasiholes. For the quasihole operators $H$ a requirement is that of mutual locality with respect to the electrons, which means that the braiding of quasiholes and electrons is trivial. This requirement implies that the operator product expansion (OPE) of the fields $H$ and $V$ is of the form
\begin{equation}
V\left(z\right)H \left(w\right) \sim \left(z-w\right)^{\ell} \tilde{H} \left(w\right)
\label{eq:e_qh_ope}
\end{equation} 
where $\ell$ is a non-negative integer and $\tilde{H}$ denotes the field resulting from the fusion of $H$ with $V$. This condition places a constraint on the possible types of quasiholes. 

A well-known example of a model wave function - that is, a trial wave function with a known parent Hamiltonian - is the Moore-Read wave function \cite{Moore1991}, which we denote by $\Psi_{\left(2,2\right)}$. Here, the notation $\Psi_{\left(n+1,k\right)}$ refers to a wave function with $n$ internal degrees of freedom and a $k$-clustering property which we refer to as a pairing property for $k=2$. For $M=0$, $\Psi_{\left(n+1,k\right)}$ has an $\su{n+1}{k}$ symmetry, while for $M>0$, this is broken down to $\su{n}{k}$. The relevant CFT for the Moore-Read model wave function is the product of the Ising CFT and the $\mathrm{u}\!\left(1\right)$ chiral boson CFT, where the correlator of the boson field $\phi$ is given by
$\left<\phi\left(z\right)\phi \left(w\right)\right> = - \log \left(z-w\right)$.
The electron and quasihole operators read
\begin{equation}
\begin{aligned}
V \left(z\right) &= \psi\left(z\right) e^{i \phi \left(z\right)}\\
H \left(w\right) &= \sigma\left(w\right) e^{\frac{i}{2} \phi\left(w\right)}
\label{eq:pf_ops}
\end{aligned}
\end{equation}
where the Majorana fermion $\psi$ and the `spin field' $\sigma$ are the primary fields of the Ising CFT and the vertex operator $e^{i\alpha \phi}$ is a primary field of the free boson CFT.  Writing $\{z\}$ for the collection $z_1,\dots,z_N$, the model wave function for the ground state is
\begin{equation}
\begin{aligned}
\Psi_{\left(2,2\right)} \left(\{z\}\right) &= \langle \psi\left(z_1\right) \cdots \psi\left(z_N\right)\rangle \langle \bg e^{i \phi\left(z_1 \right)} \cdots e^{i\phi \left(z_N\right)}\rangle \\
&= \mathrm{Pf} \left(\frac{1}{z_i-z_j}\right) \prod_{i<j} \left(z_i-z_j\right).
\label{eq:pf_bos}
\end{aligned}
\end{equation}

Because of the fusion rule $\sigma \times \sigma = 1 + \psi$ of the spin field $\sigma$, the many-quasihole wave `function' has different components labeled by a fusion channel index $p$, i.e. the specific way in which the spin fields fuse to the identity. The wave function with $2m$ quasiholes has $2^{m-1}$ components \cite{Nayak1996}, or (chiral) conformal blocks, given by
\begin{equation}
\Psi_{\left(2,2\right)}^{(p)} \left(\{w\},\{z\}\right) = \bb{\sigma\left(w_1\right) \cdots \sigma\left(w_{2m}\right)}^{(p)} \prod_{i<j} \left(z_i-z_j\right) \prod_{i,j} \left(z_i -w_j\right)^{\frac{1}{2}} \prod_{i<j}w_{ij}^{\frac{1}{4}}.
\end{equation} 
Here $w_{ij} = w_i -w_j$, the $w_i$ are assumed to be radially ordered, i.e. $|w_1|<...<|w_{2m}|$ and we have adopted the notation
\begin{equation}
\begin{aligned}
\langle \sigma (w_1) \cdots \sigma (w_{2m}) X \rangle ^{(p)} &= \langle \sigma (w_1) \cdots \sigma (w_{2m}) \psi (z_1) \cdots \psi (z_N)\rangle^{\left(p\right)}\\
\end{aligned}
\label{eq:bb_notation}
\end{equation}
with X denoting a string of Majorana fermions $\psi$.
The explicit wave functions involving arbitrarily many quasiholes and electrons for the Moore-Read wave function were found in Refs. \cite{Ardonne2010,Bonderson2011}. The implementation of the (four-) quasi-hole states as topological protected q-bits was studied Refs. \cite{dassarma2005,georgiev2006}.

The conformal blocks $\Psi_{\left(2,2\right)}^{(p)}$ transform non-trivially amongst themselves when the quasiparticle coordinates are exchanged. That is, exchanging $w_i$ and $w_j$ and analytically continuing the wave function, $\Psi_{\left(2,2\right)}^{(p)} \to  \sum_{p'} \left(U^{\left(2,2\right)}_{ij}\right)^{p}_{p'} \Psi_{\left(2,2\right)}^{(p')}$ with $U^{\left(2,2\right)}_{ij}$ a unitary braid matrix. The collection of braid matrices, which were found in Ref. \cite{Nayak1996}, forms a unitary representation of the braid group on $2m$ strands. 

\section{Paired spin-singlet states}
\label{sec:paired_spin_singlets}

\subsection{Model wave functions}
\label{sec:general_n_paired}
The Moore-Read wave function $\Psi_{\left(2,2\right)}$ is the simplest example of a paired `spin-singlet' state, denoted $\Psi_{\left(n+1,2\right)}$, which are studied in this paper. The pairing property of $\Psi_{\left(2,2\right)}$ may be verified by inspection of Eq. \eqref{eq:pf_bos}, or by considering the OPE
\begin{equation}
V\left(z\right) V\left(z'\right) \sim \left(z-z'\right)^0 e^{2i\phi \left(z\right)}
\end{equation}
between the electron operators, using $\psi\left(z\right) \psi\left(z'\right) \sim 1/(z-z')$. In fact, the Moore-Read wave function is the unique, densest zero-energy eigenstate of a certain three-body Hamiltonian \cite{greiter92,read96}. Consequently, the Moore-Read wave function may also be obtained by symmetrizing two bosonic Laughlin wave functions as observed by Cappelli et al. \cite{Cappelli1999}. Denoting the Laughlin wave functions by $\Psi_{\left(2,1\right)}$,
\begin{equation}
\begin{aligned}
\Psi_{\nsu{2}{2}} (\{z\})& = \frac{1}{\mathcal{N}} \sum_{S_1,S_2} \prod_{z\in S_1} \prod_{i<j} \left(z_i-z_j\right)^2 \prod_{z\in S_2} \prod_{i<j} \left(z_i-z_j\right)^2\\
&= \frac{1}{\mathcal{N}} \sum_{S_1,S_2} \Psi_{\nsu{2}{1}} (S_1) \Psi_{\nsu{2}{1}} (S_2) 
\label{eq:mooreread_cappelli}
\end{aligned}
\end{equation}
as Eq. \eqref{eq:mooreread_cappelli} obeys the same vanishing properties and has the same degree. Here, the coordinates $\{z\}$ are partitioned into two `layers' $S_1,S_2$  of equal size\footnote{The number of `electrons' in the ground state must be even in order for the fields $\psi$ to fuse to the identity.}, and the sum is over all inequivalent partitions. We consider two partitions to be equivalent if they are related by a layer permutation $S_1 \leftrightarrow S_2$.

The paired spin-singlet state $\Psi_{\left(n+1,2\right)}$ can be viewed as a generalization of the Moore-Read wave function to particles having $n$ internal quantum numbers. These wave functions have an underlying $\su{n+1}{2}$ symmetry. Additionally, they are also non-zero when two particles are at the same position, and vanish when three particles are brought together (quadratically when the three particles are identical, linearly otherwise).

Generalizing Eq. \eqref{eq:pf_ops} there are $n$ electron operators $V_{\alpha}$ which factor into a ``parafermion''  \cite{gepner87} $\psi_{\alpha}$ generalizing the Majorana fermion $\psi$, and a vertex operator of $n$ independent chiral bosons $\phi=\left(\phi_1,\dots,\phi_n\right)$:
\begin{equation}
V_{\alpha} \left(z\right) = \psi_{\alpha} \left(z\right) e^{iv_\alpha \cdot \phi \left(z\right)/\sqrt{2}}. 
\label{eq:pss_el}
\end{equation}
Here $\alpha=1,\dots,n$ and $v_\alpha$ is a vector: to avoid clutter in the notation, we do not write vector-superscripts.
The factor $\sqrt{2}$ in the vertex operator is included so that the vectors $v_\alpha$ are simple in terms of the roots of $\su{n+1}{}$, see
Appendix~\ref{sec:appendix_pf_n}.
In particular, they should obey $v_\alpha \cdot v_\beta = 1+\delta_{\alpha\beta}$, so that the OPE of two electron operators reads
\begin{equation}
V_{\alpha} \left(z\right) V_{\beta} \left(z'\right) \sim \left(z-z'\right)^{0} e^{i\left(v_\alpha+v_\beta\right)\cdot \phi \left(z\right)/\sqrt{2}}
\end{equation}
in accordance with the pairing property of $\Psi_{\left(n+1,2\right)}$. The paired spin-singlet states are the unique densest, zero-energy eigenstates of the same three-body Hamiltonian that has the Moore-Read state as its ground state (it is understood that the Hamiltonian treats all particle types equally).
Generalizing Eq. \eqref{eq:mooreread_cappelli}, $\Psi_{\left(n+1,2\right)}$ can be obtained by symmetrizing the following generalized Halperin wave functions
\begin{equation}
\Psi_{\left(n+1,1\right)} \left(\{z\}\right) = \prod_{\alpha=1}^{n} \prod_{i<j}^{N_\alpha} \left(z_{i}^{\alpha} - z_{j}^{\alpha}\right)^2 \prod_{\alpha<\alpha'}^{n} \prod_{i,j}  \left(z_{i}^{\alpha} - z_{j}^{\alpha'}\right).
\end{equation} 
Here $N_\alpha$ denotes the number of particles with index $\alpha$, with coordinates $z_{i}^{\alpha}$. Hence, the model wave functions $\Psi_{\left(n+1,2\right)}$ can be expressed as
\begin{equation}
\begin{aligned}
\Psi_{\left(n+1,2\right)} \left(\{z\}\right)&= \langle \prod_{\alpha=1}^{n} \prod_{i=1}^{N_\alpha} \psi_{\alpha} \left( z_i^{\alpha}\right) \rangle \, \left[\Psi_{\left(n+1,1\right)}\left(\{z\}\right)\right]^{\frac{1}{2}} \\
&= \frac{1}{\mathcal{N}} \sum_{S_1,S_2} \Psi_{\left(n+1,1\right)} \left(S_1\right) \Psi_{\left(n+1,1\right)} \left(S_2\right).
\end{aligned}
\label{eq:psi-relations}
\end{equation}
In the symmetrized representation, each layer $S_a = \{S_a^{1},S_a^{2},\ldots,S_a^{n}\}$ with $a=1,2$ contains half the coordinates with a given index $\alpha$.

The relevant CFT that describes the paired spin-singlet states is the $\su{n+1}{2}$ WZW CFT (see \cite{DiFrancesco1997} for an introduction). These CFTs can be written as a product of a parafermion theory $\pf{n+1}{2}{n}$ and $n$ free boson CFTs \cite{gepner87}, which leads to the expression Eq. \eqref{eq:pss_el} and Eq. \eqref{eq:pf_ops} for $n=1$ in which case the parafermion CFT is the Ising CFT. The more general parafermion CFTs are described in Appendix \ref{sec:appendix_pf}. The electron operators $V_\alpha$ are currents of the $\su{n+1}{2}$ WZW model, as described in Appendix \ref{sec:app_wzw_general}.

The fundamental quasiholes are represented by primary fields $H_\mu$ of the WZW model, where $\mu=0,1,\ldots,n$ labels the different types: a quasihole with a pseudospin index ($\mu=1,\ldots,n$) or a ``spinless'' quasihole ($\mu=0$). These operators read
\begin{equation}
H_{\mu} \left(w\right) = \sigma_{\mu} \left(w\right) e^{i q_\mu \cdot \phi /\sqrt{2}},
\label{eq:h_operator_n}
\end{equation}
where $\sigma_\mu$ is a spin field of the parafermion theory. In order that the operators $H_\mu$ have the correct OPEs with the electron operators,
Eq.~\eqref{eq:e_qh_ope} with $\ell=0$, the inner products have to satisfy
\begin{align}
v_{\alpha} \cdot v_{\beta} &= 1+\delta_{\alpha \beta} &
q_0 \cdot v_\alpha &= 1 &
q_\alpha \cdot v_\beta &= \delta_{\alpha \beta} \nonumber\\
q_0 \cdot q_0 &= \frac{n}{n+1} &
q_0 \cdot q_\alpha &= \frac{1}{n+1} &
q_\alpha \cdot q_{\beta} &= \delta_{\alpha \beta}-\frac{1}{n+1} \ .
\label{eq:inner_products}
\end{align} 
The quasihole wave function can be expressed as a correlator of operators $H_\mu$ and $V_\alpha$, or in terms of two copies of $\Psi_{\left(n+1,1\right)}$ with quasiholes inserted in the layers $S_1,S_2$. In particular, the operator $H_\mu$ is equivalent to the insertion a quasihole in one of the layers, which becomes a non-abelian quasihole after the symmetrization procedure.

We are mainly interested in four-quasihole wave functions. For the simplest case, where all quasiholes carry the index $\mu=1$, the two conformal blocks $(p=0,1)$ read 
\begin{equation}
\begin{aligned}
\Psi_{\left(n+1,2\right)}^{(p)} \left(\{w\},\{z\}\right) &= \bb{\sigma_1 \left(w_1\right) \sigma_1\left(w_2\right) \sigma_1\left(w_3\right) \sigma_1 \left(w_4\right)}^{(p)} \left[\Psi_{\left(n+1,1\right)}\right]^{\frac{1}{2}}\\
&\times \prod_{i=1}^{N_1} \prod_{j=1}^{4} \left(z_i^1 -w_j\right)^{\frac{1}{2}} \prod_{i<j} w_{ij}^{\frac{n}{2\left(n+1\right)}}. 
\label{eq:pss_conf_block}
\end{aligned}
\end{equation}
Here, $X$ denotes a string of parafermions $X= \prod_{\alpha,i} \psi_\alpha \left(z_i^\alpha\right)$. To express the conformal blocks in a symmetrized representation, we first define the wave functions 
\begin{equation}
\Psi_{ab;cd}(\{z\}) =\frac{1}{2\mathcal{N}} \sum_{S_1,S_2} \prod_{z\in S_1^1} (z-w_a)(z-w_b) \Psi_{\nsu{n+1}{1}}(S_1) \prod_{z\in S_2^1} (z-w_c)(z-w_d) \Psi_{\nsu{n+1}{1}}(S_2)
\label{eq:paired_qh_sym}
\end{equation}
where $S_a^1$ denotes the coordinates with index $\mu=1$ in layer $a$. Only two of the three possible symmetrized wave functions $\Psi_{12;34},\Psi_{13;24}$, and $ \Psi_{14;23}$ are linearly independent, as was seen in Ref. \cite{Nayak1996} for the case $n=1$. In particular, the wave functions $\Psi_{ab;cd}$ are related by
\begin{equation}
\left(1-x\right)\Psi_{14;23} =\Psi_{13;24} - x \Psi_{12;34}
\label{eq:paired_qh_sym_nel}
\end{equation}
in terms of the following anharmonic ratio
\begin{equation}
x = \frac{w_{12} w_{34}}{w_{13} w_{24}},\quad 1-x = \frac{w_{14}w_{23}}{w_{13}w_{24}}.
\label{eq:anharmonic}
\end{equation}
We note that the convention for the anharmonic ratio used here differs from the one used in for instance Ref. \cite{Knizhnik1984,Ardonne2007}, but agrees with the convention in Ref. \cite{Nayak1996}. The reason for picking the current convention is that the $w_i$ are properly radially ordered, namely, after an appropriate conformal transformation, we have
$w_1 = 0$, $w_2 = x$, $w_3 =1$ and $w_4 = \infty$.

The $\Psi_{ab;cd}$ obey the same vanishing properties as the conformal blocks Eq. \eqref{eq:pss_conf_block}, when either electrons or electrons and quasiholes are taken to the same point. As a result, and by virtue of Eq. \eqref{eq:paired_qh_sym_nel}, each conformal block may be expanded in the basis $\Psi_{12;34}, \Psi_{13;24}$ as
\begin{equation}
\Psi_{\left(n+1,2\right)}^{(p)} \left(\{w\},\{z\}\right)= A^{(p)} \left(\{w\}\right) \Psi_{12;34} \left(\{z\}\right)+ B^{(p)} \left(\{w\}\right) \Psi_{13;24} \left(\{z\}\right)
\label{eq:master_formula}
\end{equation}
where the expansion coefficients $A^{(p)},B^{(p)}$ depend only on the $w_i$ and ensure the correct behavior when quasiholes are brought to the same position.

\subsection{Master formulas and braiding}
\label{sec:strategy}
Following Ref. \cite{Ardonne2007}, relations like Eq. \eqref{eq:master_formula} which relate the conformal blocks to symmetrized wave functions open up the possibility of finding explicit expressions for two-quasihole and four-quasihole wave functions. In turn, this allows us to study the braiding properties of the quasiholes by finding the monodromies of the four-quasihole wave functions, i.e. the transformation properties of the conformal blocks under exchanges of quasihole positions. Such equations are therefore referred to as `master formulas'.

In the following, we obtain various master formulas for different types of quasiholes. By taking limits of the master formulas, letting the electron positions coincide with each other or with the quasihole positions, the expansion coefficients $A^{\left(p\right)},B^{\left(p\right)}$ are determined \cite{Ardonne2007}. In particular, we employ operator product expansions of the parafermions $\psi_\alpha$ and spin fields $\sigma_\mu$, found in Appendix \ref{sec:appendix_pf}, to reduce the correlator to a four-point function of spin fields. The latter can be determined using the results obtained in Ref. \cite{Knizhnik1984}, where closely related four point functions of primary fields in the $\su{n+1}{2}$ WZW CFT were found explicitly by solving the Knizhnik-Zamolodchikov equation. The spin field four point functions are presented in Appendix \ref{sec:appendix_4_point_functions}. 

Using the solutions of the coefficients $A^{\left(p\right)},B^{\left(p\right)}$ in terms of the $w_i$ we find the manifest transformation of the conformal block $\Psi_{\nsu{n+1}{2}}^{\left(p\right)}$ under $w_i\leftrightarrows w_j$:
\begin{equation}
\begin{aligned}
\Psi_{\nsu{n+1}{2}}^{\left(p\right)} &\to A'^{\left(p\right)} \Psi_{12;34}' + B'^{\left(p\right)} \Psi_{13;24}' \\
&= \sum_{p'}\left(U_{ij}^{(n+1,2)}\right)^{p}_{p'}  \Psi_{\nsu{n+1}{2}}^{(p')} .
\label{eq:paired_block_transf}
\end{aligned}
\end{equation} 
Here, $\left(U_{ij}^{(n+1,2)}\right)^p_{p'}$ is the $2\times2$ braid matrix corresponding to the given transformation. In particular, we determine the matrices corresponding to the transformations
 \begin{enumerate}
 \item $w_1\leftrightarrows w_2$, or $x \to \frac{-x}{1-x}$
 \item $w_1 \leftrightarrows w_3$, or $x \to 1-x$
 \item $w_2\leftrightarrows w_3$, or $x \to \frac{1}{x}$
 \end{enumerate}
in terms of the anharmonic ratio $x$ defined in Eq. \eqref{eq:anharmonic}. The braid matrices for the more general wave functions $\Psi^M$ (see Eq. \eqref{eq:model_wf}) are obtained afterwards and differ from the  bosonic ($M=0$) braid matrices by a global phase only. 
 
This analysis hinges on the explicit form of the four point functions of spin fields. Unfortunately the explicit form of correlators involving more than four spin fields is much harder to obtain. Therefore, although the conformal blocks and symmetrized wave functions can be written down, the expansion coefficients $A^{(p)},B^{(p)},\ldots$ can not be determined easily in the same way. Additionally we assume that the braiding statistics is determined by the manifest transformation of the wave function alone (holonomy=monodromy), i.e. that there is no additional contribution to the statistics coming from the Berry phase.

\section{Braiding for the paired $\su{4}{2}$ spin-singlet state}
\label{sec:n=3_braiding}

\subsection{The paired $\su{4}{2}$ spin-singlet state}
\label{sec:n=3_wf}
The (bosonic) model wave function $\Psi_{\left(4,2\right)}$ has the two equivalent representations
\begin{equation}
\begin{aligned}
\Psi_{\nsu{4}{2}} (\{z\}) &= \langle \prod_{i=1}^{N_{1}} \psi_1 (z_i^1) \prod_{i=1}^{N_{2}} \psi_2 (z_i^2) \prod_{i=1}^{N_3} \psi_3 (z_i^3)\rangle [\Psi_{\nsu{4}{1}}\left(\{z\}\right)]^{\frac{1}{2}} \\
&= \frac{1}{\mathcal{N}} \sum_{S_1,S_2} \Psi_{\nsu{4}{1}}(S_1) \Psi_{\nsu{4}{1}} (S_2).
\label{eq:n=3_gs}
\end{aligned}
\end{equation}
An explicit representation of the vectors $v_\alpha$ and $q_\mu$ that satisfy the correct inner products in this case are given in Eq. \eqref{eq:n=3_vecs}. The number of electrons of each pseudospin type must be even -- this ensures the parafermions fuse to the identity, or that the sets of coordinates can be partitioned into two equal sized sets in the symmetrized representation. 

The prefactor $\mathcal{N}$ may be fixed by taking pairs of parafermions to the same point, i.e. letting $z_{2j}^{\alpha} \to z_{2j-1}^{\alpha}$ for $j=1,\ldots,N_{\alpha}/2$ and $\alpha=1,2,3$. Using the OPEs of the parafermions (see Appendix \ref{sec:appendix_pf_n})
\begin{equation}
\psi_{\alpha} (z) \psi_{\alpha} (z') \sim \frac{1}{z-z'}
\label{eq:parafermion_opes}
\end{equation}  
and taking the aforementioned limit of Eq. \eqref{eq:n=3_gs}, one finds $\mathcal{N} = 2^{\frac{1}{2}(N_{1}+N_{2}+N_3)-1}$.

There are four quasihole operators: a spinless quasihole $H_0$, as well spinful quasiholes $H_1, H_2, H_3$. The simplest two-quasihole wave function is obtained by inserting two identical quasiholes $H_\mu$; the wave function reads
\begin{equation}
\begin{aligned}
\Psi_{\nsu{4}{2}} (w_1,w_2,\{z\}) &= \bb{ \sigma_\mu (w_1) \sigma_\mu (w_2) } [\Psi_{\nsu{4}{1}}]^{\frac{1}{2}} \prod_{i} (z_i^\mu -w_1)^{\frac{1}{2}} (z_i^\mu -w_2)^{\frac{1}{2}} w_{12}^{\frac{3}{8}}\\
&= \frac{A(\{w\})}{2\mathcal{N}} \sum_{S_1,S_2} \prod_{z\in S_1^{\mu}} \left(z-w_1\right)\Psi_{\nsu{4}{1}} (S_1) \prod_{z\in S_2^\mu} \left(z-w_2\right)\Psi_{\nsu{4}{1}}(S_2).
\label{eq:n=3_2qh}
\end{aligned}
\end{equation}
Here $S_a^\mu$ denotes all coordinates with pseudospin index $\mu$, where we adopt the convention $S_a^0=S_a$. Indeed, the spinless quasiholes `couple' to all types of electrons. In Eq. \eqref{eq:n=3_2qh}, the additional factor $A$ depending on $w_1,w_2$ is fixed by requiring that both sides are equal in the limit $w_2\to w_1$. Using the OPEs of the spin fields, this yields $A = w_{12}^{\frac{1}{8}}$. 

In the following sections, we present the relevant master formulas for the four-quasihole wave functions. We insert two pairs of identical quasiholes for simplicity, which leaves the cases
\begin{enumerate}[label=(\Roman*)]
\item $m_\mu=4$, corresponding to the insertions $\bb{\sigma_\mu \sigma_\mu \sigma_\mu \sigma_\mu}$ where $\mu=0,1,2,3$;
\item $m_{\mu}=m_{\mu'}=2$, for $\mu'\neq \mu$, where we consider two orderings of quasiholes corresponding to the insertions $\bb{\sigma_\mu \sigma_\mu \sigma_{\mu'} \sigma_{\mu'}}$ and $\bb{\sigma_\mu \sigma_{\mu'} \sigma_{\mu} \sigma_{\mu'} }.$
\end{enumerate}

For the bosonic wave function the different quasihole types are related by a symmetry so that their braid matrices are identical. In particular, it is enough to consider the above cases for (I) $\mu=1$ and (II) $\mu=1,\mu'=2$. The symmetry relating quasihole types is broken for $M>0$, and the braid matrices differ by an overall (global) phase from the bosonic braid matrices. The braid matrices for the wave functions with $M>0$ are presented in Section \ref{sec:n=3_ferm_wfs}.

\subsection{The case $m_{1}=4$}
\label{sec:n=3_case1}
We consider the quasihole wave function with $m_{1}=4$ and take $N_{1}=2$ and $N_{2}=N_3=6$. We label the coordinates $z_1,z_2,z_3,\ldots,z_{14}$, omitting the pseudospin index. Then, the master formula reads:
\begin{equation}
\begin{aligned}
\Psi_{\nsu{4}{2}}^{\left(p\right)} (\{w\},\{z\}) & = \bb{\sigma_{1}(w_1) \sigma_{1}(w_2)\sigma_{1}(w_3) \sigma_{1}(w_4) }^{\left(p\right)} [\Psi_{\nsu{4}{1}}]^{\frac{1}{2}}\\
&\times  \prod_{i=1}^{2} \prod_{j=1}^{4} (z_i-w_j)^{\frac{1}{2}} \prod_{i<j} w_{ij}^{\frac{3}{8}} \\
&= A^{\left(p\right)} (\{w\} ) \Psi_{12,34} + B^{\left(p\right)} (\{w\})\Psi_{13,24}.
\label{eq:n=3_master1}
\end{aligned}
\end{equation}
We then take the following three limits of Eq. \eqref{eq:n=3_master1}:
\begin{equation}
\begin{aligned}
\label{eq:n=3_limits1a}
(\mathrm{i})&: z_2\to z_1, z_4\to z_3 , z_6 \to z_5 ,\ldots, z_{14}\to z_{13}\\
(\mathrm{ii})&: z_3\to z_1,z_4\to z_2, z_6 \to z_5 ,\ldots, z_{14}\to z_{13}; z_{1}\to w_3, z_{2}\to w_4\\
(\mathrm{iii})&:z_3\to z_1, z_4\to z_2, z_6\to z_5,\ldots, z_{14}\to z_{13}; z_{1} \to w_2, z_{2}\to w_4.
\end{aligned}
\end{equation}
To obtain expressions for $A^{\left(p\right)}$ and $B^{\left(p\right)}$, only two limits are strictly necessary. The third limit is taken to fix the phases of the four-point function of spin fields: this is explained in more detail in Appendix \ref{sec:appendix_4_point_functions}. These limits reduce the full correlators to four-point functions, namely
\begin{equation}
\begin{aligned}
\bb{\sigma_1\sigma_1\sigma_1\sigma_1}^{\left(p\right)} &\overset{(\mathrm{i})}{\to} \langle \sigma_1 \sigma_1\sigma_1\sigma_1\rangle^{\left(p\right)}\\
\bb{\sigma_1 \sigma_1\sigma_1\sigma_1}^{\left(p\right)} & \overset{(\mathrm{ii})}{\to} \langle \sigma_1\sigma_1\sigma_2\sigma_2\rangle^{\left(p\right)}\\
\bb{\sigma_1 \sigma_1 \sigma_1 \sigma_1}^{\left(p\right)} &\overset{(\mathrm{iii})}{\to} \!  \langle \sigma_1 \sigma_2\sigma_1\sigma_2\rangle^{\left(p\right)}.
\label{eq:n=1_limits1b}
\end{aligned}
\end{equation}
Taking the limits of the symmetrized wave functions $\Psi_{12;34}$ and $\Psi_{13;24}$ as well, one finds the equations
\begin{eqnarray}
A^{\left(p\right)} (\{w\})+ B^{\left(p\right)} (\{w\})&=& \langle\sigma_{1}(w_1) \sigma_{1}(w_2) \sigma_{1}(w_3)\sigma_{1}(w_4)\rangle^{\left(p\right)} w_{12}^{\frac{3}{8}} w_{13}^{\frac{3}{8}} w_{14}^{\frac{3}{8}} w_{23}^{\frac{3}{8}} w_{24}^{\frac{3}{8}} w_{34}^{\frac{3}{8}}\\
B^{\left(p\right)} (\{w\}) &=&\langle \sigma_{1}(w_1) \sigma_{1}(w_2) \sigma_{2}(w_3) \sigma_{2}(w_4)\rangle ^{\left(p\right)}w_{12}^{\frac{3}{8}}w_{13}^{\frac{7}{8}}w_{14}^{-\frac{1}{8}}w_{23}^{-\frac{1}{8}}w_{24}^{\frac{7}{8}}w_{34}^{\frac{3}{8}}\\
A^{\left(p\right)} (\{w\})&=& \langle  \sigma_{1}(w_1) \sigma_{2}(w_2) \sigma_{1}(w_3) \sigma_{2}(w_4)  \rangle ^{\left(p\right)} w_{12}^{\frac{7}{8}} w_{13}^{\frac{3}{8}}w_{14}^{-\frac{1}{8}}w_{23}^{-\frac{1}{8}}w_{24}^{\frac{3}{8}}w_{34}^{\frac{7}{8}}
\label{eq:n=3_coeff1}
\end{eqnarray}
By using the four point functions of spin fields, which are determined in Appendix \ref{sec:appendix_4_point_functions}, we find
\begin{equation}
\begin{aligned}
A^{\left(p\right)} (\{w\})&=\left(-1\right)^{p}\left[w_{12}w_{34}\right]^{\frac{7}{8}}x^{-\frac{1}{8}}\left(1-x\right)^{\frac{3}{4}}h^{\frac{p}{2}}\big[\mathcal{F}_{2}^{p}\left(x\right)-\frac{x}{1-x}\mathcal{F}_{1}^{p}\left(x\right)\big]\\
B^{\left(p\right)}(\{w\})	&=\left(-1\right)^{p}\left[w_{12}w_{34}\right]^{\frac{7}{8}}x^{-\frac{1}{8}}\left(1-x\right)^{-\frac{1}{4}}h^{\frac{p}{2}}\mathcal{F}_{1}^{p}\left(x\right)
\label{eq:n=3_sol1}
\end{aligned}
\end{equation}
In Eq. \eqref{eq:n=3_sol1}, $\sqrt{h}=\frac{1}{4}2^{-\frac{1}{6}}$ (see Appendix \ref{sec:appendix_4_point_functions}) and the functions $\mathcal{F}_{1}^{p},\mathcal{F}_2^{p}$ are given in Eq. \eqref{eq:F_general}. 

\subsection{The case $m_{1}=m_{2}=2$}
\label{sec:n=3_case2}
We consider the four quasihole wave function with $m_{1}=m_{2} =2$, taking $N_{1}=N_{2}=2$ and $N_3=4$. As in the previous section, we label the coordinates $z_1,z_2,\ldots,z_8$, omitting the pseudospin indices. We consider two possible orderings of the four operators, corresponding to $\bb{\sigma_1\sigma_1 \sigma_2 \sigma_2}^{\left(p\right)}$ and $\bb{\sigma_1 \sigma_2 \sigma_1 \sigma_2}^{\left(p\right)}$. 

Note that in this case, there are only two natural ways of dividing the quasiholes over the two layers in the Cappelli representation. For the first case, the symmetrized wave functions are $\Psi_{13;24}$ and $\Psi_{14;23}$. For the second case, they are $\Psi_{12;34}$ and $\Psi_{14;23}$. 

\subsubsection*{First case}
For the first case, we write 
\begin{equation}
\begin{aligned}
\Psi_{\nsu{4}{2}}^{\left(p\right)} (\{w\},\{z\})&=\bb{\sigma_{1}(w_1) \sigma_{1}(w_2) \sigma_{2}(w_3)\sigma_{2}(w_4)}^{\left(p\right)}  [\Psi_{\nsu{4}{1}} ]^{\frac{1}{2}} \\
&\times \prod_{i,j=1}^{2} (z_i-w_j)^{\frac{1}{2}} \prod_{i,j=3}^{4} (z_i-w_j)^{\frac{1}{2}}  w_{12}^{\frac{3}{8}} w_{13}^{-\frac{1}{8}} w_{14}^{-\frac{1}{8}} w_{23}^{-\frac{1}{8}}w_{24}^{-\frac{1}{8}} w_{34}^{\frac{3}{8}}\\
&= A_1^{\left(p\right)} (\{w\})\Psi_{13;24} + B_1^{\left(p\right)} (\{w\})\Psi_{14;23}.
\label{eq:n=1_master2a}
\end{aligned}
\end{equation}
We then consider the limits 
\begin{equation}
\begin{aligned}
(\mathrm{i})&: z_2\to z_1, z_4\to z_3 , z_6\to z_5, z_8 \to z_7\\
(\mathrm{ii})&: z_3\to z_1, z_4 \to z_2, z_6\to z_5, z_8 \to z_7;  z_1 \to w_3, z_2 \to w_4,
\label{eq:n=3_limits2a}
\end{aligned}
\end{equation}
which give the equations
\begin{equation}
\begin{aligned}
A_{1}^{\left(p\right)}(\{w\})+B_1^{p}(\{w\})&=\left(-1\right)^{p}\left[w_{12}w_{34}\right]^{-\frac{1}{8}}x^{\frac{7}{8}}\left(1-x\right)^{-\frac{1}{4}}h^{\frac{p}{2}}\mathcal{F}_{1}^{p}\left(x\right)\\
A_1^{\left(p\right)}(\{w\})+(1-x)B_1^{p}(\{w\})&=\left(-1\right)^{p}\left[w_{12}w_{34}\right]^{-\frac{1}{8}}x^{\frac{7}{8}}\left(1-x\right)^{\frac{3}{4}}h^{\frac{p}{2}}\left[\mathcal{F}_{1}^{p}\left(x\right)+\mathcal{F}_{2}^{p}\left(x\right)\right].
\label{eq:n=3_coeff2a}
\end{aligned}
\end{equation}
As in the previous section, we use the four point functions of the spin fields to find
\begin{equation}
\begin{aligned}
A_{1}^{\left(p\right)}(\{w\})&=\left(-1\right)^{p}\left[w_{12}w_{34}\right]^{-\frac{1}{8}}x^{-\frac{1}{8}}\left(1-x\right)^{\frac{3}{4}}h^{\frac{p}{2}}\mathcal{F}_{2}^{p} \left(x\right)\\
B_{1}^{\left(p\right)}(\{w\})&=\left(-1\right)^{p}\left[w_{12}w_{34}\right]^{-\frac{1}{8}}x^{-\frac{1}{8}}\left(1-x\right)^{-\frac{1}{4}}h^{\frac{p}{2}}\left[x\mathcal{F}_{1}^{p}\left(x\right)-\left(1-x\right)\mathcal{F}_{2}^{p}\left(x\right)\right].
\label{eq:n=3_sol2a}
\end{aligned}
\end{equation}

\subsubsection*{Second case}
For the second case, the master formula reads
\begin{equation}
\begin{aligned}
\Psi_{\nsu{4}{2}}^{\left(p\right)} (\{w\},\{z\}) &= \bb{\sigma_{1}(w_1) \sigma_{2}(w_2)\sigma_{1}(w_3)\sigma_{2}(w_4) }^{\left(p\right)} \big[ \Psi_{\nsu{4}{1}}\big]^{\frac{1}{2}} \prod_{i=1,2}(z_{i}-w_{1})^{\frac{1}{2}}(z_{i}-w_{3})^{\frac{1}{2}}\\ 
&\times\prod_{j=3,4}(z_{j}-w_{2})^{\frac{1}{2}}(z_{j}-w_{4})^{\frac{1}{2}} w_{12}^{-\frac{1}{8}}w_{13}^{\frac{3}{8}}w_{14}^{-\frac{1}{8}}w_{23}^{-\frac{1}{8}}w_{24}^{\frac{3}{8}}w_{34}^{-\frac{1}{8}}\\
&= A_2^{\left(p\right)} (\{w\}) \Psi_{12;34} + B_2^{\left(p\right)} (\{w\}) \Psi_{14;32}.
\label{eq:n=3_master2b}
\end{aligned}
\end{equation}
In this case, we take the limits
\begin{equation}
\begin{aligned}
(\mathrm{iii})&:\, z_2\to z_1, z_4\to z_3, z_6 \to z_5, z_8\to z_7\\
(\mathrm{iv})&:\,z_3 \to z_1, z_4\to z_2, z_6 \to z_5, z_8 \to z_7; z_1 \to w_2, z_2\to w_4.
\label{eq:n=3_limits2b}
\end{aligned}
\end{equation}
which give the solutions
\begin{equation}
\begin{aligned}
A_{2}^{\left(p\right)}(\{w\})	&=\left(-1\right)^{p}\left[w_{12}w_{34}\right]^{-\frac{1}{8}}x^{-\frac{1}{8}}\left(1-x\right)^{\frac{3}{4}} h^{\frac{p}{2}}\mathcal{F}_{2}^{p}\left(x\right)\\
B_{2}^{\left(p\right)}(\{w\})&=-\left(-1\right)^{p}\left[w_{12}w_{34}\right]^{-\frac{1}{8}}x^{\frac{7}{8}}\left(1-x\right)^{-\frac{1}{4}}h^{\frac{p}{2}}\mathcal{F}_{1}^{p}\left(x\right).
\label{eq:n=3_sol2b}
\end{aligned}
\end{equation}

\subsection{The case $m_{0}=4$}
\label{sec:n=3_case3}
We take $m_{0}=4$ and $N_{1}=N_{2}=N_3=2$. The conformal blocks have the expressions
\begin{equation}
\begin{aligned}
\Psi_{\nsu{4}{2}}^{\left(p\right)}(\{w\},\{z\}) &= \bb{\sigma_0 (w_1) \sigma_0 (w_2) \sigma_0 (w_3) \sigma_0 (w_4)}^{\left(p\right)} [\Psi_{\nsu{4}{1}}]^{\frac{1}{2}} \prod_{i,j} (z_i-w_j)^{\frac{1}{2}} \prod_{i<j} w_{ij}^{\frac{3}{8}}\\
&= A^{\left(p\right)}(\{w\}) \Psi_{12;34} + B^{\left(p\right)} (\{w\}) \Psi_{13;24}.
\label{eq:n=3_master3}
\end{aligned}
\end{equation}
We take the limits
\begin{equation}
\begin{aligned}
(\mathrm{i})&:\, z_2\to z_1, z_4\to z_3, z_6 \to z_5\\
(\mathrm{ii})&:\, z_4\to z_3, z_6\to z_5; z_1 \to w_3, z_2\to w_4
\label{eq:n=3_limits3}
\end{aligned}
\end{equation}
which reduce the four point functions to $\bb{\sigma_0 \sigma_0 \sigma_0 \sigma_0}^{\left(p\right)} \to \langle \sigma_0 \sigma_0 \sigma_0 \sigma_0\rangle^{\left(p\right)}$ for $\left(\mathrm{i}\right)$ and $\bb{\sigma_0 \sigma_0 \sigma_0 \sigma_0}^{\left(p\right)} \to \langle \sigma_0 \sigma_0 \sigma_1 \sigma_1 \rangle ^{\left(p\right)}$ for limit $\left(\mathrm{ii}\right)$. This yields the same equations as in Eq. \eqref{eq:n=3_sol1}, i.e. the braid matrices for spinless quasiholes in the bosonic case are be the same as the braid matrices for the spinful quasiholes. This result was to be expected: it follows from the $\su{4}{}$ symmetry which is unbroken in the case $M=0$.

\subsection{Braiding transformations}
\label{sec:n=3_braids}
By keeping track of how the coefficients $A^{\left(p\right)},B^{\left(p\right)}$ and the symmetrized quasihole wave functions $\Psi_{ab;cd}$ transform, we find the manifest transformation of the conformal blocks $\Psi_{\left(4,2\right)}^{(p)}$. The transformations of the $\Psi_{ab;cd}$ are obtained straightforwardly, using Eq. \eqref{eq:paired_qh_sym_nel}. The transformations of the coefficients $A^{\left(p\right)}$, $B^{\left(p\right)}$ follow from the transformations of the anharmonic ratios and the transformations of the functions $\mathcal{F}_{i}^{p}$ which are presented in Appendix \ref{sec:appendix_F_transf}. The combined transformation yields 
\begin{equation}
\Psi_{\nsu{4}{2}}^{\left(p\right)}\to \sum_{p'}\left(U_{ij}^{(4,2)}\right)^{p}_{p'} \Psi_{\nsu{4}{2}}^{(p')},
\label{eq:n=3_block_transf}
\end{equation}
where $(U^{(4,2)}_{ij})^p_{p'}$ is the $2\times2$ braid matrix corresponding to the transformation $w_i \leftrightarrows w_j$. 
We now present the matrices corresponding to the transformations $w_1 \leftrightarrows w_2$, $w_1 \leftrightarrows w_3$ and $w_2 \leftrightarrows w_3$, found from the solutions of the master formulas in Sections \ref{sec:n=3_case1} - \ref{sec:n=3_case3}. The braid matrices read
\begin{eqnarray}
\label{eq:n=3_braid12}
U_{12}^{(4,2)}	&=&\left(-1\right)^{\frac{1}{8}}\left(\begin{array}{cc}
1 & 0\\
 0 & \left(-1\right)^{\frac{2}{3}}
\end{array}\right)\\
U_{23}^{(4,2)}	&=&\frac{\left(-1\right)^{\frac{5}{8}}}{\sqrt{3}}\left(\begin{array}{cc}
1 & -\left(-1\right)^{\frac{1}{3}}\sqrt{2}\\
-\left(-1\right)^{\frac{1}{3}}\sqrt{2} & \left(-1\right)^{-\frac{1}{3}}
\end{array}\right) \\
\label{eq:n=3_braid13}
U_{13}^{(4,2)}	&=&\frac{\left(-1\right)^{\frac{7}{8}}}{\sqrt{3}}\left(\begin{array}{cc}
1 & \sqrt{2}\\
\sqrt{2} & -1
\end{array}\right) .
\label{eq:n=3_braid23}
\end{eqnarray}
All of the matrices are found for the case $m_{1}=4$, while the case $m_{1}=m_{2}=2$ yields only the matrices $U_{12}^{(4,2)}$ and $U_{13}^{(4,2)}$, one for each ordering. As was mentioned before, the symmetry between the quasiholes in the bosonic case means these are the correct braid matrices in the general cases $m_{\mu} =4$ and $m_{\mu}=m_{\mu'}=2$ as well. 

The matrices $U_{ij}$ constitute a two-dimensional representation of the braid group, as they are unitary and satisfy the `Yang-Baxter' relation $U_{13}=U_{12}U_{23}U_{12} = U_{23}U_{12}U_{23}$. Moreover, the braid matrices are closely related to the braid matrices associated with quasiholes in the $k=4$ Read-Rezayi wave function \cite{Read1999}, which we write $\Psi_{\left(2,4\right)}$. The braid matrices for the Read-Rezayi states are given in Ref. \cite{Ardonne2007,Slingerland2001}. The close relation is due to the rank-level duality between the WZW CFTs $\su{4}{2}$ for the paired spin-singlet and $\su{2}{4}$ for the Read-Rezayi state \cite{Naculich1990}. Denoting the braid matrices for $\Psi_{\left(n+1,k\right)}$ by $U_{ij}^{\left(n+1,k\right)}$, the matrices satisfy
\begin{equation}
\begin{aligned}
U_{12}^{\left(4,2\right)}  &= \left(-1\right)^{1/24} \,\overline{U}_{12}^{\left(2,4\right)}\\
U_{23}^{\left(4,2\right)} &= \left(-1\right)^{1/24}\,\overline{U}_{23}^{\left(2,4\right)} \\
U_{13}^{\left(4,2\right)} &= \left(-1\right)^{1/8}\,\,\,\overline{U}_{13}^{\left(2,4\right)} ,
\end{aligned}
\end{equation}
where the overline indicates that the rows and columns (i.e., the order of the fusion channels) of the matrix are swapped.
This is explained in more detail in Appendix~\ref{app:rank-level-duality}.
\subsection{Wave functions for general $M$}
\label{sec:n=3_ferm_wfs}
To obtain the braid matrices for the $M>0$ wave functions $\Psi_{\left(4,2\right)}^{M}$, we modify the electron and quasihole operators for the $M=0$ case. First, we adopt the following representation \cite{Tournois2017} of the vectors $v$ and $q$
\begin{equation}
\begin{aligned}
v_1 &= \left(\frac{2}{\sqrt{3}},0,-\frac{2}{\sqrt{6}}\right),v_2=\left(\frac{2}{\sqrt{3}},\frac{1}{\sqrt{2}},\frac{1}{\sqrt{6}}\right), v_3 = \left(\frac{2}{\sqrt{3}},-\frac{1}{\sqrt{2}},\frac{1}{\sqrt{6}}\right)\\
q_1 &= \left(\frac{1}{2\sqrt{3}},0,-\frac{2}{\sqrt{6}}\right),q_2=\left(\frac{1}{2\sqrt{3}},\frac{1}{\sqrt{2}},\frac{1}{\sqrt{6}}\right), q_3 = \left(\frac{1}{2\sqrt{3}},-\frac{1}{\sqrt{2}},\frac{1}{\sqrt{6}}\right), q_0 = \left(\frac{3}{2\sqrt{3}},0,0\right)
\label{eq:n=3_vecs}
\end{aligned}
\end{equation}
which satisfy the inner products Eq.~\eqref{eq:inner_products}. These vectors ensure charge neutrality in all sectors except the first, so that $\bg$ depends on the field $\phi_1$ only, which describes charge. To obtain the model wave function $\Psi_{\left(4,2\right)}^{M}$ without quasiholes, we change the first components to
\begin{equation}
v_{\alpha}^1 \to v_{\alpha,M}^1 = \sqrt{\frac{4+6M}{3}}
\label{eq:n=3_vecsM1}
\end{equation}
which yields the appropriate modification to the wave function as in Eq. \eqref{eq:model_wf}. Introducing quasiholes, the appropriate change to $q_\mu^1$ follows from the requirement of mutual locality, so that the inner products between the vectors $q_\mu$ and $v_\alpha$ is unchanged for $M>0$. For the vectors $q$, this yields
\begin{equation}
\begin{aligned}
q_{\alpha,M} \cdot q_{\alpha',M} &= q_{\alpha} \cdot q_{\alpha'} - \frac{M}{2(4+6M)} \\
q_{0,M} \cdot q_{\alpha,M} &= q_{0} \cdot q_{\alpha} - \frac{3M}{2(4+6M)}\\
q_{0,M} \cdot q_{0,M} &= q_0 \cdot q_0 - \frac{9M}{2(4+6M)}.
\label{eq:n=3_vecsM2}
\end{aligned}
\end{equation}

The new conformal blocks $\Psi_{\nsu{4}{2}}^{M\,\left(p\right)}$ then differ from their bosonic $M=0$ counterparts by the full Jastrow factor $\prod_{i<j} \left(z_i-z_j\right)^M$ and similar factors of the quasihole coordinates which we denote by $\Xi$:
\begin{equation}
\Psi_{\nsu{4}{2}}^{M\left(p\right)} \left(\{w\},\{z\}\right) =  \Xi(\{w\})\Psi_{\nsu{4}{2}}^{\left(p\right)} \left(\{w\},\{z\}\right)\prod_{i<j} \left(z_i-z_j\right)^M .
\label{eq:n=3_ferm_wf}
\end{equation}
The factors $\Xi$ for the different cases are
\begin{itemize}
\item $m_{\alpha} = 4$ and $m_{\alpha} = m_{\alpha'} = 2$: $\Xi = \prod_{i<j} w_{ij}^{-\frac{M}{4\left(4+6M\right)}}$
\item $m_{0} = 4$: $\Xi = \prod_{i<j} w_{ij}^{-\frac{9M}{4\left(4+6M\right)}}$
\end{itemize} 

The factor $\Xi$ leads to additional, global phases of the braid matrices, while the Jastrow factor has no effect. In particular, they break the symmetry between the braiding behavior of the different quasiholes. The resulting braid matrices for the quasiholes with spin are
\begin{equation}
\begin{aligned}
U_{12}^{(4,2),M}	&= \left(-1\right)^{-\frac{M}{4(4+6M)}} U_{12}^{(4,2)}\\
U_{23}^{(4,2),M} &= \left(-1\right)^{-\frac{M}{4(4+6M)}} U_{23}^{(4,2)}\\
U_{13}^{(4,2),M}	&=\left(-1\right)^{-\frac{3M}{4(4+6M)}}U_{13}^{(4,2)} .
\end{aligned}
\end{equation}
The updated braid matrices for the spinless quasiholes read
\begin{equation}
\begin{aligned}
U_{12}^{(4,2),M}	&=\left(-1\right)^{-\frac{9M}{4(4+6M)}}U_{12}^{(4,2)}\\
U_{23}^{(4,2),M}	&=\left(-1\right)^{-\frac{9M}{4(4+6M)}}U_{23}^{(4,2)}\\
U_{13}^{(4,2),M}	&=\left(-1\right)^{-\frac{27M}{4(4+6M)}}U_{13}^{(4,2)} .
\end{aligned}
\end{equation}

\section{Braiding for the paired $\su{n+1}{2}$ spin-singlet states}
\label{sec:r_braiding}
We turn to the braiding of the fundamental quasiholes in the paired $\su{n+1}{2}$ spin-singlet state Eq. \eqref{eq:r_gs}. There are $n$ electron operators and $n+1$ quasihole operators $H_\mu$, given by Eq. \eqref{eq:h_operator_n}, in terms of parafermions $\psi_\alpha$ and spin fields $\sigma_\mu$ of the parafermion theory $\pf{n+1}{2}{n}$. The model wave function for the ground state reads
\begin{equation}
\begin{aligned}
\Psi_{\nsu{n+1}{2}} (\{z\})&= \langle\prod_{\alpha=1}^{n} \prod_{i=1}^{N_{\alpha}} \psi_{\alpha} (z_{i}^{\alpha}) \rangle \big[\Psi_{\nsu{n+1}{1}} \big]^{\frac{1}{2}}\\
&=\frac{1}{\mathcal{N}} \sum_{S_1,S_2} \Psi_{\nsu{n+1}{1}} (S_1) \Psi_{\nsu{n+1}{1}} (S_2),
\label{eq:r_gs}
\end{aligned}
\end{equation}
where the OPE of the parafermion fields fixes the normalization $\mathcal{N}=2^{\frac{1}{2} (\sum_\alpha N_{\alpha}) -1 }$. We consider the master formula for the case $m_1=4$, Eq. \eqref{eq:master_formula}, since it yields all braid matrices. The master formula is:
\begin{equation}
\begin{aligned}
\Psi_{\nsu{n+1}{2}}^{\left(p\right)} (\{w\},\!\{z\}) &= \bb{\sigma_{1}(w_1) \sigma_{1}(w_2) \sigma_{1}(w_3)\sigma_{1}(w_4)}^{\left(p\right)} \big[\Psi_{\nsu{n+1}{1}}\big]^{\frac{1}{2}} \prod_{i,j} (z_{i}^{1}-w_j)^{\frac{1}{2}}\prod_{i<j} w_{ij}^{\frac{n}{2(n+1)}}\\
&= A^p (\{w\}) \Psi_{12;34} + B^p (\{w\}) \Psi_{13;24}.
\label{eq:r_master}
\end{aligned}
\end{equation}
We consider the simplest case where $N_{1}=2$ and $N_{i\geq 2}=6$, taking the three limits
\begin{equation}
\begin{aligned}
\label{eq:r_limits}
(\mathrm{i})&:\, z_{2k}\to z_{2k-1} \quad \text{for}\,\,\, k\geq1 \\
(\mathrm{ii})&:\, z_{3}\to z_{1}, z_{4} \to z_{2}, z_{2k}\to z_{2k-1}\,\, \text{for}\,\,\, k\geq 3; z_1 \to w_{3}, z_{2}\to w_{4}\\
(\mathrm{iii})&:\, z_{3}\to z_{1}, z_{4}\to z_{2}, z_{2k}\to z_{2k-1}\,\, \text{for}\,\,\, k\geq 3; z_1 \to w_{2}, z_{2}\to w_{4}.
\end{aligned}
\end{equation}
The four-point functions of spin fields are presented in Appendix \ref{sec:appendix_4_point_functions}, and yield the solutions
\begin{equation}
\begin{aligned}
A^{p}(\{w\})	&=\left(-1\right)^p[w_{12}w_{34}]^{\frac{2n}{n+1}-2\Delta}x^{2\Delta-\frac{n}{n+1}}(1-x)^{\frac{n}{n+1}}h^{\frac{p}{2}} [\mathcal{F}_{2}^{p}\left(x\right)-\frac{x}{1-x}\mathcal{F}_1^p \left(x\right)]  \\
B^{p}(\{w\})	&=\left(-1\right)^p[w_{12}w_{34}]^{\frac{2n}{n+1}-2\Delta}x^{2\Delta-\frac{n}{n+1}}(1-x)^{-\frac{1}{n+1}}h^{\frac{p}{2}}\mathcal{F}_{1}^{p}\left(x\right).
\label{eq:r_coeff}
\end{aligned}
\end{equation}
Here $2\Delta=\frac{n(n+2)}{(n+1)(n+3)}$. Further, $\sqrt{h}$ and the functions $\mathcal{F}_i^p$ are given in Appendix \ref{sec:appendix_pf_n}. Using the transformations of the functions $\mathcal{F}_{i}^p$ presented in Appendix \ref{sec:appendix_F_transf}, it is straightforward to obtain the braid matrices
\begin{eqnarray}
\label{eq:r_braid12}
U_{12}^{(n+1,2)}	&=&\left(-1\right)^{\frac{n}{(n+1)(n+3)}} \left(\begin{array}{cc}
1 & 0\\
0 & \left(-1\right)^{\frac{n+1}{n+3}} 
\end{array}\right)\\
\label{eq:r_braid23}
U_{23}^{(n+1,2)}	&=&\frac{\left(-1\right)^{2\Delta}}{d_n}\left(\begin{array}{cc}
1 & -\left(-1\right)^{\frac{2}{n+3}}\sqrt{d_{n}^{2}-1}\\
-\left(-1\right)^{\frac{2}{n+3}}\sqrt{d_{n}^{2}-1} & -\left(-1\right)^{\frac{4}{n+3}}
\end{array}\right)\\
\label{eq:r_braid13}
U_{13}^{(n+1,2)}	&=&\frac{\left(-1\right)^{\frac{n(n+4)}{(n+1)(n+3)}}}{d_n}\left(\begin{array}{cc}
1 & \sqrt{d_{n}^{2}-1}\\
\sqrt{d_{n}^{2}-1} & -1
\end{array}\right) .
\end{eqnarray}
In these expressions, $d_{n}=2\cos\left(\frac{\pi}{n+3}\right)$, see Appendix \ref{sec:appendix_F_transf}. For $n=3$, they reduce to the matrices in Eqs. \eqref{eq:n=3_braid12}-\eqref{eq:n=3_braid23}. For $n=2$, these braid matrices agree with the results obtained in Ref. \cite{Ardonne2007} for the NASS case\footnote{In comparing the matrices in Eqs \eqref{eq:r_braid12}-\eqref{eq:r_braid23} for $n=2$ to the braid matrices obtained for the NASS case, one should be aware that the anharmonic ratios differ. The braid matrices for $n=2$ and the matrices listed in  Ref. \cite{Ardonne2007} are related by a similarity transformation determined by $U_{12}=U_{34}$.}, while for $n=1$ they agree with the braid matrices for the Moore-Read wave function \cite{Nayak1996}. Again, the matrices $U_{ij}$ constitute a unitary representation of the braid group, i.e. they satisfy $U_{13}=U_{12} U_{23} U_{12} = U_{23}U_{12}U_{23}$. Finally, the matrices are closely related to the braid matrices (see \cite{Ardonne2007}) of the Read-Rezayi $\Psi_{\left(2,n+1\right)}$ states, see Appendix~\ref{app:rank-level-duality} for more detail.

Generalizing the discussion in Section \ref{sec:n=3_ferm_wfs}, the braid matrices for the wave functions for general $M$ read
\begin{equation}
\begin{aligned}
U_{12}^{\left(n+1,2\right),M}&=\left(-1\right)^{\frac{-M}{(n+1)(n+1+2nM)}}U_{12}^{\left(n+1,2\right)}\\
U_{23}^{\left(n+1,2\right),M} &=\left(-1\right)^{\frac{-M}{(n+1)(n+1+2nM)}}U_{23}^{\left(n+1,2\right)}\\
U_{13}^{\left(n+1,2\right),M} &=\left(-1\right)^{\frac{-3M}{(n+1)(n+1+2nM)}}U_{13}^{\left(n+1,2\right)}.
\label{eq:braid_rM_spin}
\end{aligned}
\end{equation}
for the spinful quasiholes, and
\begin{equation}
\begin{aligned}
U_{12}^{\left(n+1,2\right),M}&=\left(-1\right)^{\frac{-n^2M}{(n+1)(n+1+2nM)}}U_{12}^{\left(n+1,2\right)}\\
U_{23}^{\left(n+1,2\right),M} &=\left(-1\right)^{\frac{-n^2M}{(n+1)(n+1+2nM)}}U_{23}^{\left(n+1,2\right)}\\
U_{13}^{\left(n+1,2\right),M} &=\left(-1\right)^{\frac{-3n^2M}{(n+1)(n+1+2nM)}}U_{13}^{\left(n+1,2\right)}.
\label{eq:braid_rM_spinless}
\end{aligned}
\end{equation}
for the spinless quasiholes.

\section{Application to Hermanns hierarchy states}
\label{sec:application}
We apply the results obtained to a series of recently introduced trial wave functions \cite{Hermanns2010} which are obtained from a hierarchy picture of successive condensation of non-abelian quasiparticles. We refer to these wave functions as \NAH{} states. They can be thought of as symmetrized copies of composite fermion (CF) \cite{Jain1989} wave functions and were studied numerically in \cite{Sreejith2011}.  In Ref. \cite{Tournois2017}, the \NAH{} states were given a CFT description by using their close relation to the paired spin-singlet states. Referring to Refs. \cite{Hermanns2010,Tournois2017} for more details, the (bosonic) \NAH{} wave functions read
\begin{equation}
\Psi_{\mathrm{Her};n} \left(\{z\}\right) = \frac{1}{\mathcal{N}} \sum_{S_1,S_2} \Psi_{\mathrm{CF};n} \left(S_1\right) \Psi_{\mathrm{CF};n} \left(S_2\right),
\label{eq:NAH}
\end{equation}
where the symmetrization is similar to that in the paired spin-singlet case except one now symmetrizes over two (bosonic) CF wave functions instead of the $\Psi_{\left(n+1,1\right)}$, with
\begin{equation}
\Psi_{\mathrm{CF};n} (\{z\})= \mathcal{S} [ \prod_{\lambda=1}^{n} \partial_{\lambda}^{\lambda-1} \prod_{\lambda=1}^{n}\prod_{i<j} (z_i^\lambda -z_j^{\lambda})^2 \prod_{\lambda<\lambda'}^{n} \prod_{i,j} (z_i^\lambda -z_j^{\lambda'})].
\label{eq:CF_bosonic}
\end{equation}
Here $\lambda$ labels the effective $\Lambda$-levels, $\lambda=1,\ldots,n$, and $\partial_\lambda^m\equiv \prod_i \frac{\partial^m}{(\partial z_i^\lambda)^m} $ is a product over derivatives of coordinates in level $\lambda$. The $\Psi_{\mathrm{CF};n}$ have $\nu = \frac{n}{n+1}$, and their fermionic counterparts constitute the positive Jain series with $\nu=\frac{n}{2n+1}$. Therefore, the bosonic \NAH{} wave function has $\nu=\frac{2n}{n+1}$ and the corresponding fermionic wave function has filling fraction $\nu=\frac{2n}{3n+1}$. The case $n=1$ corresponds to the Moore-Read state, while for $n=2,3$, the \NAH{} wave functions are, after particle-hole conjugation, candidates for $\nu=2+\frac{3}{7}$ and $\nu=2+\frac{2}{5}$, respectively.

To see the relation between the \NAH{} wave functions and the paired spin-singlet states we recognize Eq. \eqref{eq:CF_bosonic} as a pseudospin symmetrization of the generalized Halperin state $\Psi_{\nsu{n+1}{1}}$, i.e.
$\Psi_{\mathrm{CF};n}(\{z\}) = \mathcal{S} [ \prod_{\alpha=1}^{n} \partial_{\alpha}^{\alpha-1} \Psi_{\nsu{n+1}{1}}\left(\{z\}\right)]$, identifying the internal quantum numbers $\alpha$ with the $\lambda$ levels. Therefore Eq. \eqref{eq:NAH} can be rewritten by doing the symmetrization over the layer first, so that
\begin{equation}
\Psi_{\mathrm{Her};n}\left(\{z\}\right)= \mathcal{S} [\prod_{\alpha=1}^{n} \partial_{\alpha}^{\alpha-1} \Psi_{\nsu{n+1}{2}}(\{z\})]
\label{eq:her_pss}
\end{equation}
in terms of the paired spin-singlet states. In Eq. \eqref{eq:her_pss}, the particles have definite pseudospin indices in the paired spin-singlet state $\Psi_{\left(n+1,2\right)}$ and the symmetrization is a sum over the ways of assigning pseudospin to the particles. 

Similarly, the quasihole model wave functions in the \NAH{} are obtained by symmetrizing paired spin-singlet states with quasiholes. The latter have $n+1$ distinct fundamental quasiholes, i.e. quasiholes with a definite pseudospin index $\mu=1,\dots,n$ or the spinless quasihole with $\mu=0$.  Although we perform a symmetrization, effectively removing internal quantum numbers, the \NAH{} wave function still has $n+1$ distinct fundamental quasiholes, discernible by the short distance behavior of the many-quasihole wave function. Considering a single quasihole for simplicity\footnote{To ensure charge neutrality we assume another quasihole is placed at infinity.}, a model wave function for a quasihole with the smallest charge is
\begin{equation}
\Psi_{\mathrm{Her};n} \left(w,\{z\}\right) = \mathcal{S}[ \prod_{\alpha} \partial_{\alpha}^{\alpha-1} \Psi_{\left(n+1,2\right)} \big(w^{\left(1\right)},\{z\}\big)],
\label{eq:her_qh}
\end{equation} 
where $w^{\left(1\right)}$ denotes a quasihole with pseudospin $\mu=1$. The model wave functions for different choices of $\mu=1,\ldots,n$ are expected to differ slightly because of the derivatives, but to have the same topological properties. The other type of fundamental quasihole corresponds to the spinless $\mu=0$ quasihole in the paired spin-singlet state. It is straightforward to generalize this to several quasiholes. 

With the quasihole wave functions in place, we now argue that the braid properties of the quasiholes in the \NAH{} are the same as those of the paired spin-singlet states studied in this paper. In writing the \NAH{} wave functions one has to perform two symmetrizations: one over identical layers as in Eq.~\eqref{eq:NAH}, and one over pseudospin as in the CF wave functions.

The symmetrization over identical layers changes the statistics of the quasiholes in the individual layers: the most famous example is the Cappelli et al. construction \cite{Cappelli1999} of the Moore-Read state, via the symmetrization of two Laughlin states. This symmetrization reduces the dimension of the Hilbert space of quasihole states, which effectively renders the quasiholes of the Laughlin layers non-abelian. Likewise, this symmetrization procedure renders the quasiholes in the generalized Halperin states non-abelian, resulting in the braiding properties of the state $\Psi_{\left(n+1,2\right)}$. 

Contrarily, it has been argued that the pseudospin symmetrization does not change the statistics of the quasiholes \cite{hansson-quasi-electron}, in accordance with the result that one can determine the statistics for the quasiholes of the CF wave functions from the K-matrix formalism \cite{wen-zee-92,wen-review}. In symmetrizing over the pseudospin, a reduction of the dimension of the Hilbert space is expected not to occur. The difference with the layer case is that the parts of the wave functions associated with different pseudospin are not identical. Although this is not a proof, it is likely that even for the \NAH{} wave functions the symmetrization over pseudospin does not alter the statistics of the quasiholes. Assuming this argument to be correct, the fundamental quasiholes in the \NAH{}{} come in two types, whose braid matrices are given by Eqs.~\eqref{eq:braid_rM_spin} and \eqref{eq:braid_rM_spinless} respectively. 

In particular, the (non-abelian) braid behavior of the fundamental quasiholes in the \NAH{} wave functions at $\nu=\frac{2n}{3n+1}$ is the same, up to an overall phase, as that of the quasiholes in the $\mathbb{Z}_{n+1}$ Read-Rezayi wave functions at $\nu=\frac{n+1}{n+3}$. For $n=2$, the \NAH{} wave function is a trial wave function for $\nu=2+\frac{3}{7}$ (after particle-hole conjugation), and one expects the quasiholes to obey $\mathbb{Z}_3$ statistics. For $n=3$, the \NAH{} wave function has the same filling factor as the $\mathbb{Z}_3$ Read-Rezayi wave function: both are wave functions for $\nu=2+\frac{2}{5}$ (after particle-hole conjugation). Interestingly, one expects $\mathbb{Z}_4$-type braiding in the \NAH{} case, which is non-universal for topological quantum computing \cite{freedman2002}, as opposed to the $\mathbb{Z}_3$ braiding expected in the Read-Rezayi case. Additionally, this differs from the Ising (Moore-Read) statistics expected for quasiholes in the Bonderson-Slingerland \cite{bonderson08} hierarchy state at $\nu=2+\frac{2}{5}$.

\section{Conclusion}
In this paper we have studied the braiding properties of the fundamental quasiholes in the paired spin-singlet states by finding explicit expressions of the quasihole wave functions and obtaining their monodromies. As expected on the basis of rank-level duality, we have shown that the non-abelian braiding properties of the quasiholes in the paired spin-singlet states are closely related to the quasiholes in the Read-Rezayi series, with the only difference an overall phase.  The extension to \emph{clustered} spin-singlet states $\Psi_{\left(n+1,k\right)}$ with $k>2$ is straightforward, although additional subtleties such as fusion multiplicities will arise, and is left to future work. Additionally, we have argued that the braid behavior of quasiholes in certain (spin polarized) non-abelian hierarchy states should agree with that of the quasiholes in the paired spin-singlet states, and have observed that if the former are the appropriate model wave functions, the expected braid properties are $\mathbb{Z}_3$-type braiding for $\nu=2+\frac{3}{7}$ and $\mathbb{Z}_4$-type braiding for $\nu=2+\frac{2}{5}$. The latter is to be contrasted with the $\mathbb{Z}_3$-type braiding based on the Read-Rezayi wave function and Ising statistics ($\mathbb{Z}_2$-type braiding) based on the state in the Bonderson-Slingerland hierarchy. 

In finding the quasihole braiding properties from the CFT wave functions, we have assumed that `holonomy=monodromy', i.e. that no additional Berry phase contributes to the braid statistics. Additionally, we have argued that the braiding properties of the paired spin-singlet states are unchanged by a symmetrization procedure. A promising method to address these matters is the matrix product state implementation of \cite{Zaletel2012}, by means of which the full Berry holonomy may be calculated numerically for large system sizes.

\subsubsection*{Acknowledgements}
We would like to thank M. Hermanns and T. H. Hansson for insightful discussions and comments on the manuscript. EA thanks K.~Schoutens for the collaboration on \cite{Ardonne2007}. This research was sponsored, in part, by the Deutsche Forschungsgemeinschaft under grant no. CRC183 (YT) and the Swedish Research Council (EA).

\begin{appendix}

\section{WZW models and current algebra}
\label{sec:app_wzw_general}
We provide details on the WZW models that underpin the paired spin-singlet states. We discuss the current algebra as well as the WZW primary fields with respect to this current algebra. Introducing a vertex representation of the currents as well as the WZW primary fields, we explicitly identify the electron and quasihole operators used to write down the model wave functions. We refer to \cite{Knizhnik1984,Fuchs,DiFrancesco1997} for more information.

\subsection{Current algebra}
The $\su{n+1}{k}$ WZW model is characterized by its current algebra, a set of OPEs of currents $J^a$ corresponding to the generators $t^a$ of $\su{n+1}{}$
\begin{equation}
J^{a}\left(z\right)J^{b}\left(w\right)\sim\frac{\frac{k}{2}\delta^{ab}}{\left(z-w\right)^{2}}+\frac{if^{abc}J^{c}\left(w\right)}{z-w},
\end{equation}
with $f^{abc}$ the structure constants, i.e. $[t^a,t^b]=i f^{abc} t^c$. 

As a simple example, we consider $n=1$. The currents $J^{1},J^{2}$ and $J^{3}$ obey the above OPEs with $f^{abc}=\epsilon^{abc}$. Alternatively one may introduce raising and lowering operators through $J^{\pm}=J^{1}\pm iJ^{2}$, so that 
\begin{equation}
\begin{aligned}
J^{+}\left(z\right)J^{-}\left(w\right)	&\sim\frac{k}{\left(z-w\right)^{2}}+\frac{2J^{3}\left(w\right)}{z-w}\\
J^{3}\left(z\right)J^{\pm}\left(w\right)	&\sim\frac{\pm J^{\pm}\left(w\right)}{z-w}\\
J^{3}\left(z\right)J^{3}\left(w\right)	&\sim\frac{\frac{k}{2}}{\left(z-w\right)^{2}}.
\end{aligned}
\end{equation}
The current algebra of the $\su{n+1}{k}$ model is generated by $J_{\alpha}^{\pm}$ with $\alpha=1,\dots,n$ which form $\su{2}{}$ subalgebras with $J^{3}_{\alpha}$. 

Following Gepner \cite{gepner87}, the vertex representation of the $\su{n+1}{k}$ current algebra is an explicit representation of the currents $J^a$ in terms of free bosons $\phi=\left(\phi_{1},\dots,\phi_{n}\right)$ and parafermions: \begin{equation}
\begin{aligned}
J_{\alpha}^{+}\left(z\right)	&=\sqrt{k}\psi_{\alpha}^{\dagger}\left(z\right)e^{iv_{\alpha}\cdot\phi/\sqrt{k}}\\
J_{\alpha}^{-}\left(z\right)	&=\sqrt{k}\psi_{\alpha}\left(z\right)e^{-iv_{\alpha}\cdot \phi/\sqrt{k}}\\
J_{\alpha}^{3}\left(z\right)	&=\frac{i\sqrt{k}}{2}v_{\alpha} \cdot \partial\phi\left(z\right).\\
\end{aligned}
\end{equation}
The vectors $v_\alpha$ obey $v_{\alpha}\cdot v_{\alpha} = 2$ and $v_{\alpha}\cdot v_\beta = 1$ if $\alpha\neq\beta$; they correspond to specific roots in the root lattice of $\su{n+1}{}$. 
It is straightforward to show that these currents generate the $\su{n+1}{k}$ current algebra, by using the OPEs
\begin{equation}
\begin{aligned}
\partial\phi_{i}\left(z\right)\partial\phi_{j}\left(w\right)&\sim\frac{-\delta_{ij}}{\left(z-w\right)^{2}}\\
\psi_{\alpha}^{\dagger}\left(z\right)\psi_{\alpha}\left(w\right)&\sim\left(z-w\right)^{-2+\frac{2}{k}} ,
\end{aligned}
\end{equation}
as well as the OPE between vertex operators
\begin{equation}
e^{iv_{\alpha}\cdot \phi\left(z\right)/\sqrt{k}}e^{iv_{\beta}\cdot\phi\left(w\right)/\sqrt{k}}\sim\left(z-w\right)^{v_{\alpha}\cdot v_{\beta}/k}e^{iv_{\alpha}\cdot \phi\left(z\right)/\sqrt{k}+iv_{\beta}\cdot\phi\left(w\right)/\sqrt{k}}. 
\end{equation}
 For $k=1$ the parafermions $\psi_{\alpha}$ are trivial, while for $k=2$ they satisfy $\psi_{\alpha}^{\dagger}=\psi_{\alpha}$. The parafermions are discussed in more detail in Appendix \ref{sec:appendix_pf}. The connection to the paired spin-singlet states is the identification of the electron operators with the raising operators:
 \begin{equation}
 V_{\alpha} \left(z\right) = J_{\alpha}^+ \left(z\right) = \sqrt{2} \psi_{\alpha} \left(z\right) e^{iv_\alpha \cdot \phi /\sqrt{2}}.
 \end{equation}

\subsection{WZW primary fields}
The primary fields in the WZW model are fields that correspond to a specific representation of the algebra $\su{n+1}{k}$. The number of irreducible representations is finite, as opposed to the algebras $\su{n+1}{}$. In particular, the representations $\hat{\Lambda}$ of $\su{n+1}{k}$ are denoted 
\begin{equation}
\hat{\Lambda}= \left(\Lambda_0;\Lambda_1,\dots,\Lambda_n\right),
\end{equation} 
where $\sum_{\mu}\Lambda_{\mu}=k$ and the $\Lambda_{\mu}$ are positive integers. Each representation $\hat{\Lambda}$ corresponds to a representation $\Lambda=\left(\Lambda_1,\dots,\Lambda_n\right)=\sum_i \Lambda_i \omega_i$ of $\su{n+1}{}$ [here $\omega_i$ are the fundamental weights]. Therefore, the possible representations $\hat{\Lambda}$ of $\su{n+1}{k}$ can be represented by the $\Lambda$ labels alone. We adopt this convention in the following, but it should be kept in mind that the proper labels carry an additional label $\Lambda_0 = k-\sum_i \Lambda_i \geq 0$.

To each representation $\Lambda$ (strictly speaking $\hat{\Lambda}$) of $\su{n+1}{k}$ corresponds a collection of fields $G^{\Lambda}$. The ``components'' correspond to the weights $\lambda$ in the representation $\Lambda$, and are denoted
$G^\Lambda_\lambda$. Thus, the field $G^\Lambda$ can be thought of as a vector of size $\dim \Lambda$ (note that we only consider one chiral half of the theory). The field $G^\Lambda$ satisfies the OPE
\begin{equation}
J^{a}\left(z\right)G^{\Lambda}\left(w\right)\sim\frac{-t_{\Lambda}^{a}G^{\Lambda}\left(w\right)}{z-w}
\end{equation}
with respect to the currents $J^a$, where $t_{\Lambda}^{a}$ is the generator $t^{a}$ in the representation $\Lambda$. 

As a simple example, the representations of $\su{2}{2}$ are $\left(2;0\right),\left(1;1\right)$ and $\left(0;2\right)$. The associated $\Lambda$ labels correspond to the trivial representation $\Lambda=0$, the fundamental representation $\Lambda=1$, and the adjoint representation $\Lambda=2$ of $\su{2}{}$. The primary fields corresponding to these representations are $G^0 = \left(G^0_0\right), G^1=\left(G^1_1,G^1_{-1}\right)$, and $G^2 = \left(G^2_2,G^2_{0},G^2_{-2}\right)$. 

In the general case, the weights $\lambda=\left(\lambda_1,\dots,\lambda_n\right)$ in the representation $\Lambda$ are obtained by subtracting simple roots
\begin{equation}
\alpha_{1} =\left(2,-1,0,\dots,0\right), \alpha_{2} = \left(-1,2,-1,0\dots,0\right), \dots , \alpha_{n}=\left(0,0,\dots,-1,2\right)
\label{eq:simple-roots}
\end{equation} 
from $\Lambda$ (see e.g. \cite{Fuchs,DiFrancesco1997}).

The WZW primary fields $G^\Lambda_\lambda$ can also be represented in terms of primary fields $\Phi^{\Lambda}_\lambda$ in the corresponding parafermion CFT:
\begin{equation}
G_{\lambda}^{\Lambda}\left(w\right)=\Phi_{\lambda}^{\Lambda}\left(w\right)e^{i\lambda \cdot\phi\left(w\right)/\sqrt{k}}.
\label{eq:pfdef}
\end{equation}
The parafermion CFTs are discussed in Appendix \ref{sec:appendix_pf}. The quasihole operators may be identified with particular WZW primary fields; the corresponding primary fields are the spin fields $\sigma_\mu$.

\section{The $\pf{n+1}{2}{n}$  parafermion CFT}
\label{sec:appendix_pf}

We provide the details on the $\pf{n+1}{2}{n}$ parafermion CFTs, see \cite{gepner87} for more information.  For the case $n=3$ we explicitly list all primary fields and their conformal dimensions, as well as the fusion rules. We also list the relevant OPE coefficients. For general $n$, we list only those details needed to perform the braiding calculation. 

\subsection{General properties}
\label{subsec:pf_general}

The $\pf{n+1}{2}{n}$ parafermion CFTs were introduced in \cite{gepner87}.
To completely specify the cosets $\pf{n+1}{2}{n}$, we have to specify the radii (or the number of primary fields) of the compactified boson theories.
These radii are $2 i (i+1)$, where $i=1,2,\ldots,n$.
The primary fields are related to the $\su{n+1}{k}$ WZW models through the relation Eq.~\eqref{eq:pfdef}. The primary fields in the parafermion CFT are of the form $\Phi^{\Lambda}_{\lambda}$, where $\Lambda$ denotes an $\su{n+1}{k}$ representation and $\lambda$ is a weight in that representation. To such primary fields, the following field identifications must be applied:
\begin{equation}
\Phi^{\Lambda}_{\lambda} = \Phi^{\Lambda}_{\lambda+k \alpha}
\label{eq:field_id1}
\end{equation}
where $\alpha$ is an element of the root lattice $Q=\mathbb{Z}\alpha_1 + \dots + \mathbb{Z}\alpha_n$, as well as
\begin{equation}
\Phi^{\left(\Lambda_1,...,\Lambda_n\right)}_{\left(\lambda_1,...,\lambda_n\right)} = \Phi^{\left(\Lambda_0,\Lambda_1...,\Lambda_{n-1}\right)}_{\left(\lambda_1+k,\lambda_2...,\lambda_n\right)}
\label{eq:field_id2}
\end{equation}
where $\Lambda_0=k-\sum_i \Lambda_i$. 

The parafermion CFT corresponding to the $\su{2}{2}$ model, which is the Ising CFT $\pf{2}{2}{}$, has the primary fields $\Phi^{0}_{0}, \Phi^{1}_{1}, \Phi^{1}_{-1}, \Phi^{2}_{2},\Phi^{2}_{0}$ and $\Phi^{2}_{-2}$ prior to field identifications. One then identifies $\Phi^{2}_{-2} \sim \Phi^{2}_{2}$ via Eq.~
\eqref{eq:field_id1} (the simple root is $\alpha=2$ in this case) and $\Phi^{2}_{2} \sim \Phi^{0}_{0},\Phi^{1}_{-1}\sim \Phi^{1}_{1}$ via Eq.~\eqref{eq:field_id2}. We are left with the three well-known primary fields $1=\Phi^{0}_{0}, \sigma = \Phi^{1}_{1}$ and $\psi=\Phi^{2}_{0}$ of the Ising CFT. 

The conformal dimensions of the primary fields follow from Eq.~\eqref{eq:pfdef} and the conformal dimensions of the WZW primary fields \cite{gepner87}; in the following we simply list the results. The braiding calculation further relies on the precise operator product expansions between primary fields, which are used to take limits of the master formulas. In general, the OPE between primary fields $\phi_i$ with conformal dimension $\Delta_i$ reads
\begin{equation}
\phi_a (z) \phi_b (w) \sim \sum_{c} C_{ab}^{c} (z-w)^{\Delta_c -\Delta_a-\Delta_b}\phi_c(w).
\label{eq:general_ope}
\end{equation}
Here $C_{ab}^{c}$ denotes an OPE coefficient, and it is non-zero only if $\phi_c$ appears in the fusion between $\phi_a$ and $\phi_b$. For $k=2$, each primary field fuses with itself to the identity, so that $C_{aa}^{1}=1$ for all $a$. For the remaining OPEs, we use the general expression for a three-point function of conformal fields:
\begin{equation}
\label{eq:general_3_point_function}
\langle\phi_{a} (z_1) \phi_{b} (z_2) \phi_c (z_3)\rangle = \frac{C_{abc}}{z_{12}^{\Delta_{a}+\Delta_{b}-\Delta_c} z_{13}^{\Delta_{a}+\Delta_c - \Delta_b} z_{23}^{\Delta_b + \Delta_c -\Delta_a}}
\end{equation}
with structure constants $C_{abc}=C_{ab}^{c} = C_{ac}^{b} = C_{bc}^{a}$. These can be determined by performing contractions of the ground state and the quasihole wave functions. 

\subsection{Details on $\pf{4}{2}{3}$}
We provide the details for the $\pf{4}{2}{3}$ parafermion CFT. The representations of the $\su{4}{2}$ algebra are
\begin{equation}
\label{eq:n=3_Lambdas}
\Lambda =
\left(0,0,0\right),\left(2,0,0\right),\left(0,2,0\right),\left(0,0,2\right),
\left(1,0,0\right),\left(0,1,0\right), \left(0,0,1\right), \left(1,1,0\right),\left(1,0,1\right),\left(0,1,1\right).
\end{equation}
The weights in the representations $\Lambda$ are found by subtracting the appropriate simple roots
\begin{equation}
\label{eq:n=3_noots}
\alpha_1 = \left(2,-1,0\right),\, \alpha_2 = \left(-1,2,-1\right),\, \alpha_3 =\left(0,-1,2\right),
\end{equation}
which leads to a list of fields $\Phi^{\Lambda}_{\lambda}$. Applying the field identifications Eqs.~\eqref{eq:field_id1},\eqref{eq:field_id2}, the $\pf{4}{2}{3}$ parafermion CFT has the following twenty primary fields
\begin{equation}
\begin{aligned}
1 & =\Phi_{0}^{0} & \psi_{1} & =\Phi_{\left(2,-1,0\right)}^{0} & \psi_{2} & =\Phi_{\left(1,1,-1\right)}^{0} & \psi_{3} & =\Phi_{\left(1,0,1\right)}^{0}\\
\psi_{12} & =\Phi_{\left(-1,2,-1\right)}^{0} & \psi_{13} & =\Phi_{\left(-1,1,1\right)}^{0} & \psi_{23} & =\Phi_{\left(0,-1,2\right)}^{0} & \psi_{123} & =\Phi_{\left(0,2,0\right)}^{0}\\
\sigma_{0} & =\Phi_{\left(1,0,0\right)}^{\left(1,0,0\right)} & \sigma_{1} & =\Phi_{\left(1,-1,0\right)}^{\left(0,0,1\right)} & \sigma_{2} & =\Phi_{\left(0,1,-1\right)}^{\left(0,0,1\right)} & \sigma_{3} & =\Phi_{\left(0,0,1\right)}^{\left(0,0,1\right)}\\
\tau_{0}&=\Phi^{\left(0,1,1\right)}_{\left(1,0,0\right)} & \tau_{1}&=\Phi^{\left(1,1,0\right)}_{\left(1,-1,0\right)} &  \tau_{2}&=\Phi^{\left(1,1,0\right)}_{\left(0,1,-1\right)} &  \tau_{3} & =\Phi^{\left(1,1,0\right)}_{\left(0,0,1\right)}\\
\rho & =\Phi_{\left(0,0,0\right)}^{\left(1,0,1\right)} & \gamma_{1} & = \Phi^{\left(1,0,1\right)}_{\left(2,-1,0\right)} & \gamma_{2} & =\Phi^{\left(1,0,1\right)}_{\left(1,1,-1\right)} & \gamma_{3} & =\Phi^{\left(1,0,1\right)}_{\left(1,0,1\right)}
\label{eq:pf_n=3_primaries}
\end{aligned} 
\end{equation}
Here we have used the shorthand $0=(0,0,0)$. The most important fields are the parafermions $\psi_1,\psi_2,\psi_3$ and the spin fields $\sigma_0,\sigma_1,\sigma_2,\sigma_3$, which are used to define the electron and fundamental quasihole operators in Eqs.~\eqref{eq:pss_el} and \eqref{eq:h_operator_n}
for $n=3$. Note that the $\lambda$ labels of the $\psi_\alpha$ are
\begin{equation}
\begin{aligned}
\left(2,-1,0\right) &= \alpha_1, \, \left(1,1,-1\right)=\alpha_1+\alpha_2, \, \left(1,0,1\right) = \alpha_1 +\alpha_2+\alpha_3
\end{aligned}
\end{equation}
in terms of the simple roots Eq.~\eqref{eq:n=3_noots}. We denote these vectors by $v_1 =\alpha_1, v_2 = \alpha_1+\alpha_2$ and $v_3=\alpha_1+\alpha_2+\alpha_3$, so that 
\begin{equation}
V_{\alpha} \left(z\right) = \Phi^{0}_{v_\alpha} \left(z\right) e^{iv_\alpha \cdot \phi\left(z\right)/\sqrt{2}}.
\end{equation}
 Similarly, denoting the $\lambda$ labels of the spin fields by $q_0 = \omega_1, q_1 =\omega_1-\omega_2, q_2 = \omega_2 -\omega_3$ and $q_3=\omega_3$, we obtain
 \begin{equation}
 \begin{aligned}
 H_0 \left(w\right) &= \Phi^{\omega_1}_{q_0}\left(w\right) e^{iq_0 \cdot \phi\left(w\right) /\sqrt{2}}\\
 H_{\alpha} \left(w\right) &= \Phi^{\omega_3}_{q_\alpha} \left(w\right) e^{iq_\alpha \cdot \phi \left(w\right) /\sqrt{2}}.
 \end{aligned}
 \end{equation}
 These vectors $v_\alpha$ and $q_\mu$ obey the correct inner products, where the inner product should be taken with respect to the quadratic form matrix of $\su{4}{2}$. 

The conformal dimensions of the primary fields are
\begin{equation}
\Delta_{\psi} = \frac{1}{2},\,\, \Delta_{\psi_{123}}=1, \,\, \Delta_{\sigma} =\frac{1}{8}, \, \,\Delta_{\tau} =\frac{5}{8}, \, \,\Delta_{\gamma}  = \frac{1}{6},\, \,\Delta_{\rho} = \frac{2}{3}.
\label{eq:n=3_conf_dims}
\end{equation}

\subsubsection{Fusion rules and OPEs}
\label{app:fusion-ope}
We list the full set of fusion rules between the primary fields in Eq. \eqref{eq:pf_n=3_primaries}. In general, the fusion rules read 
\begin{equation}
\Phi^{\Lambda}_{\lambda} \times \Phi^{\Lambda}_{\lambda} = \sum_{\Lambda'' \in \Lambda \times \Lambda' }\Phi^{\Lambda''}_{\lambda+\lambda'}
\label{eq:pf-fusion}
\end{equation}
where $\Lambda\times\Lambda'$ denotes the fusion of the representations $\Lambda\times\Lambda'$, which may be obtained by the Littlewood-Richardson rule. Note that the field identifications Eqs.~\eqref{eq:field_id1},\eqref{eq:field_id2} may need to be used on the fusion outcomes.
 
The parafermions have simple, abelian fusion rules: they have $\Lambda = 0$, so their $\lambda$ labels add modulo $2Q$ by virtue of Eq.~\eqref{eq:pf-fusion}.
We reminder the reader that $Q$ is the root lattice, and $k = 2$ in this case.
The fusion table is  \\
\\
\begin{minipage}{\linewidth}
\centering
\begin{tabular}{c|ccccccc}
 & $\psi_{1}$  & $\psi_{2}$  & $\psi_{3}$ & $\psi_{123}$ & $\psi_{23}$ & $\psi_{13}$ & $\psi_{12}$\tabularnewline
 \hline
$\psi_1$ & $1$ & $$ & $$ & $$ & $$  &$$ & $$ \tabularnewline
$\psi_2$ &$\psi_{12}$ &$1$ & & & & &\tabularnewline
$\psi_3$ &$\psi_{13}$ &$\psi_{23}$ & $1$& & & &  \tabularnewline
$\psi_{123}$ &$\psi_{23}$ &$\psi_{13}$ & $\psi_{12}$& $1$& & & \tabularnewline
$\psi_{23}$ &$\psi_{123}$ &$\psi_{3}$ & $\psi_{2}$& $\psi_{1}$&$1$ & &  \tabularnewline
$\psi_{13}$ &$\psi_3$ &$\psi_{123}$ & $\psi_{1}$& $\psi_{2}$&$\psi_{12}$ &$1$ & \tabularnewline
$\psi_{12}$ &$\psi_2$ &$\psi_{1}$ & $\psi_{123}$& $\psi_{3}$&$\psi_{13}$ &$\psi_{23}$ &$1$\tabularnewline
\end{tabular}
\end{minipage}
\\
\\
\\
For the remaining fusion rules, we first note the following:
\begin{equation}
\sigma_\mu \times \psi_{123} = \tau_{\mu}.
\label{eq:sigma_tau}
\end{equation}
By associativity of the fusion rules, the $\sigma$ fusion rules encode all $\tau$ fusion rules as well. Then, the following fusion tables encode all fusion rules:\\
\\
\vspace{0.5cm}
\begin{minipage}{.55\linewidth}
\centering
\begin{tabular}{c|ccccccc}
 & $\psi_{1}$  & $\psi_{2}$  & $\psi_{3}$ & $\psi_{123}$ & $\psi_{23}$ & $\psi_{13}$ & $\psi_{12}$\tabularnewline
 \hline
$\sigma_{0}$ & $\sigma_{1}$ & $\sigma_{2}$ & $\sigma_{3}$ & $\tau_{0}$ & $\tau_{1}$ & $\tau_{2}$ & $\tau_{3}$\tabularnewline
$\sigma_{1}$ & $\sigma_{0}$ & $\tau_{3}$ & $\tau_{2}$ & $\tau_{1}$ & $\tau_{0}$ & $\sigma_{3}$ & $\sigma_{2}$\tabularnewline
$\sigma_{2}$ & $\tau_{3}$ & $\sigma_{0}$ & $\tau_{1}$ & $\tau_{2}$ & $\sigma_{3}$ & $\tau_{0}$ & $\sigma_{1}$\tabularnewline
$\sigma_{3}$ & $\tau_{2}$ & $\tau_{1}$ & $\sigma_{0}$ & $\tau_{3}$ & $\sigma_{2}$ & $\sigma_{1}$ & $\tau_{0}$\tabularnewline
\end{tabular}
\end{minipage}
\begin{minipage}{.55\linewidth}
\centering
\begin{tabular}{c|ccccccc}
 & $\psi_{1}$  & $\psi_{2}$  & $\psi_{3}$ & $\psi_{123}$ & $\psi_{23}$ & $\psi_{13}$ & $\psi_{12}$\tabularnewline
 \hline
$\rho$ & $\gamma_{1}$ & $\gamma_{2}$ & $\gamma_{3}$ & $\rho$ & $\gamma_{1}$ & $\gamma_{2}$ & $\gamma_{3}$\tabularnewline
$\gamma_{1}$ & $\rho$ & $\gamma_{3}$ & $\gamma_{2}$ & $\gamma_{1}$ & $\rho$ & $\gamma_{3}$ & $\gamma_{2}$\tabularnewline
$\gamma_{2}$ & $\gamma_{3}$ & $\rho$ & $\gamma_{1}$ & $\gamma_{2}$ & $\gamma_{3}$ & $\rho$ & $\gamma_{1}$\tabularnewline
$\gamma_{3}$ & $\gamma_{2}$ & $\gamma_{1}$ & $\rho$ & $\gamma_{3}$ & $\gamma_{2}$ & $\gamma_{1}$ & $\rho$\tabularnewline
\end{tabular}
\end{minipage}
\vspace{0.5cm}
\begin{minipage}{.55\linewidth}
\centering
\begin{tabular}{c|cccc} 
 & $\sigma_{0}$ & $\sigma_{1}$ & $\sigma_{2}$ & $\sigma_{3}$\tabularnewline
\hline 
$\sigma_{0}$ & $1+\rho$ &  &  & \tabularnewline
$\sigma_{1}$ & $\psi_{1}+\gamma_{1}$ & $1+\rho$ &  & \tabularnewline
$\sigma_{2}$ & $\psi_{2}+\gamma_{2}$ & $\psi_{12}+\gamma_{3}$ & $1+\rho$ & \tabularnewline
$\sigma_{3}$ & $\psi_{3}+\gamma_{3}$ & $\psi_{13}+\gamma_{2}$ & $\psi_{23}+\gamma_{1}$ & $1+\rho$
\tabularnewline
\end{tabular}
\end{minipage}
\vspace{0.1cm}
\begin{minipage}{.54\linewidth}
\centering
\begin{tabular}{c|cccc} 
 & $\sigma_{0}$ & $\sigma_{1}$ & $\sigma_{2}$ & $\sigma_{3}$\tabularnewline
\hline 
$\rho$ & $\sigma_0+\tau_0$ & $\sigma_1+ \tau_1$  &$\sigma_2+\tau_2$  &$\sigma_3+\tau_3$ \tabularnewline
$\gamma_{1}$ & $\sigma_1+\tau_1$ & $\sigma_0+\tau_0$ &$\sigma_3+\tau_3$  &$\sigma_2+\tau_2$ \tabularnewline
$\gamma_{2}$ & $\sigma_2+\tau_2$ & $\sigma_{3}+\tau_{3}$ & $\sigma_0+\tau_0$ & $\sigma_{1}+\tau_{1}$ \tabularnewline
$\gamma_{3}$ & $\sigma_3+\tau_3$ & $\sigma_2+\tau_2$ & $\sigma_1+\tau_1$ & $\sigma_0+\tau_0$
\tabularnewline
\end{tabular}
\end{minipage}
\vspace{0.5cm}
\begin{minipage}{\linewidth}
\centering
\begin{tabular}{c|cccc} 
 & $\rho$ & $\gamma_{1}$ & $\gamma_{2}$ & $\gamma_{3}$\tabularnewline
\hline 
$\rho$ & $1+\psi_{123}+\rho$ & & &  \tabularnewline
$\gamma_{1}$ &$\psi_1 + \psi_{23} + \gamma_{1}$ & $1+\psi_{123}+\rho$ & &  \tabularnewline
$\gamma_{2}$ & $\psi_{2}+\psi_{13}+\gamma_{2}$ & $\psi_{3}+\psi_{12}+\gamma_3$ & $1+\psi_{123}+\rho$ &  \tabularnewline
$\gamma_{3}$ & $\psi_{3}+\psi_{12}+\gamma_3$ & $\psi_{2}+\psi_{13}+\gamma_2$ & $\psi_{1}+\psi_{23}+\gamma_1$ & $1+\psi_{123}+\rho$
\tabularnewline
\end{tabular}
\end{minipage}

In this particular CFT, the fusion rule Eq. \eqref{eq:sigma_tau} implies that the braiding properties of the fields $\tau$ are closely related to those of the fields $\sigma$. In particular the difference is a sign: braiding two $\tau$ fields is equivalent to braiding a pair of $\sigma$ and $\psi_{123}$ around another pair, which is seen to give a relative minus sign compared to the braiding of the $\sigma$ fields alone. The corresponding WZW primary fields $T_\mu = \tau_\mu e^{iq_\mu \phi /\sqrt{2}}$ yield the same braid matrices as the $H_\mu$, again up to a sign. We have verified this by explicitly calculating the $F$ and $R$ symbols for the representations $\Lambda=\left(0,1,1\right)$ and $\Lambda=\left(1,1,0\right)$ to which the $\tau$ fields correspond, using the quantum group approach \cite{Slingerland2001}.

We turn to the coefficients appearing in the operator product expansions of the field. By performing contractions of the ground state and quasihole wave functions, we reduce the correlators to three point functions, which determines several OPE coefficients. The coefficients for parafermions read:
\begin{equation}
\begin{aligned}
\label{eq:ope_coeff1}
C_{\psi_1 \psi_2 \psi_{12}} = C_{\psi_1 \psi_3 \psi_{13}} = C_{\psi_2 \psi_3 \psi_{23}} = C_{\psi_{12} \psi_{13}\psi_{23}} &= \frac{1}{\sqrt{2}}\\
C_{\psi_{1}\psi_{23} \psi_{123}} =C_{\psi_2 \psi_{13} \psi_{123}} = C_{\psi_3 \psi_{12} \psi_{123}} &= 1 .\\
\end{aligned}
\end{equation}
The structure of the remaining relevant OPE coefficients is
\begin{equation}
\begin{aligned}
\label{eq:ope_coeff2}
C_{\sigma \sigma' \psi}&= C_{\sigma \tau \psi_{123}} = \frac{1}{\sqrt{2}}, C_{\sigma \tau \psi} = 1,\\
C_{\sigma \sigma \rho} &= \sqrt{3\sqrt{h}}, C_{\sigma \sigma' \gamma} = \sqrt{-4\sqrt{h}}.
\end{aligned}
\end{equation}

\subsubsection*{The sector $\rho$}
\label{sec:rho}
The weight $\left(0,0,0\right)$ in the adjoint representation $\Lambda=\left(1,0,1\right)$ has multiplicity three - this means that the field $\rho=\Phi^{\left(1,0,1\right)}_{\left(0,0,0\right)}$ actually consists of three independent Virasoro primary fields. A similar feature was noted in in the NASS case \cite{Ardonne2007}, where the equivalent sector splits up into two independent Virasoro primary fields. We proceed in a similar way as in that paper, defining fields $\rho_{\mu}$ by
\begin{equation}
\sigma_{\mu} \left(w\right)\sigma_{\mu}\left(w'\right) \sim \left(w-w'\right)^{-2\Delta_\sigma} + \left(w-w'\right)^{\Delta_\rho-2 \Delta_\sigma} \sqrt{3\sqrt{h}} \rho_\mu\left(w'\right).
\label{eq:rho_def}
\end{equation}
This distinction between the sector $\rho$ and the fields $\rho_\mu$ is necessary to ensure consistency of the four-point functions of spin fields: studying their behavior also leads to the choice of OPE coefficient $\sqrt{3\sqrt{h}}$ above - see Appendix \ref{sec:appendix_4_point_functions}. Additionally one finds the OPEs
\begin{equation}
C_{\rho_\mu \rho_{\mu'}}^{1} = -\frac{1}{3},
\label{eq:ope_coeff}
\end{equation}
i.e. the fields $\rho_\mu$ are not independent. They may be written in terms of the three independent fields $\rho_c, \rho_s, \rho_t$ as 
\begin{equation}
\begin{aligned}
\rho_0 &= -\rho_c\\
 \rho_1 &= \frac{1}{3} \rho_c + 0 \rho_s - \frac{2\sqrt{2}}{3} \rho_t\\
 \rho_2 &= \frac{1}{3} \rho_c + \sqrt{\frac{2}{3}} \rho_s + \frac{\sqrt{2}}{3}\rho_t\\
 \rho_3 &= \frac{1}{3} \rho_c - \sqrt{\frac{2}{3}} \rho_s + \frac{\sqrt{2}}{3}\rho_t.
\end{aligned}
\end{equation}

\subsection{Details on $\pf{n+1}{2}{n}$ }
\label{sec:appendix_pf_n}

We provide the details on the CFT $\pf{n+1}{2}{n}$ needed to perform the braiding calculation. The primary fields are labeled by the representations $\hat{\Lambda} = (\Lambda_0;\Lambda_1,...,\Lambda_n)$ with $\sum_{\mu} \Lambda_\mu = 2$, and weights $\lambda$ obtained by subtracting the simple roots Eq.~\eqref{eq:simple-roots}. The important fields after the field identifications are
\begin{equation}
\begin{aligned}
\rho &=  \Phi^{\alpha_1+\alpha_n}_{0} & \psi_{1}&=\Phi^{0}_{\alpha_1} & \psi_{2} &= \Phi^{0}_{\alpha_1+\alpha_2} & \cdots & & \psi_{n}&=\Phi^{0}_{\alpha_1+\cdots+\alpha_n}\\
\psi_{12} &= \Phi^{0}_{\alpha_2} &  \psi_{23} &= \Phi^{0}_{\alpha_3} & \psi_{34} &=\Phi^{0}_{\alpha_4} & \cdots & & \psi_{(n-1)n}&=\Phi^{0}_{\alpha_n}\\
\sigma_0 &= \Phi^{\omega_1}_{q_0} & \sigma_{1}&=\Phi^{\omega_{n}}_{q_1} & \sigma_{2} &= \Phi^{\omega_n}_{q_2} & \cdots & &  \sigma_{n}&=\Phi^{\omega_n}_{q_n}
\label{eq:pf_n_primaries}
\end{aligned} 
\end{equation}
This table is not exhaustive: there are many more primary fields within the CFT, but in order to perform the calculation of the braid behavior of the fundamental quasiholes we only need detailed knowledge of the fields listed above. The electron operators are 
\begin{equation}
V_{\alpha} \left(z\right) = \Phi^{0}_{v_\alpha} \left(z\right) e^{iv_\alpha \cdot \phi \left(z\right) /\sqrt{2}}
\end{equation}
with $v_1 = \alpha_1, v_2 = \alpha_1 +\alpha_2, \dots , v_n = \alpha_1+\dots+\alpha_n$ and the quasihole operators read
\begin{equation}
\begin{aligned}
H_{0} \left(w\right) &= \Phi^{\omega_1}_{q_0} \left(w\right) e^{iq_0 \cdot \phi \left(w\right)/\sqrt{2}}\\
H_{\alpha}\left(w\right) &= \Phi^{\omega_n}_{q_\alpha} \left(w\right) e^{iq_\alpha \cdot \phi \left(w\right)/\sqrt{2}}
\end{aligned}
\end{equation}
with $q_0 =\omega_1$, $q_1 = \omega_1-\omega_2, q_2 = \omega_2-\omega_3, \dots, q_n = \omega_n$. The conformal dimensions of these fields are \cite{gepner87} 
\begin{equation}
\label{eq:pf_n_conf_dims}
\Delta_{\psi}=\frac{1}{2},\,\, \Delta_{\sigma}=\frac{n}{4(n+3)},\, \,\Delta_\rho=\frac{n+1}{n+3}.
\end{equation}
Using Eq.~\eqref{eq:pf-fusion} and the field identifications Eq. \eqref{eq:field_id1} the relevant fusion rules are
\begin{equation}
\begin{aligned}
\psi_{\alpha} \times \psi_{\alpha} &= 1\\
\psi_{1}\times \psi_{2} &= \psi_{12}, \psi_{1} \times \psi_{3}=\psi_{13},\dots \\
\psi_{\alpha(\alpha+1)}\times \sigma_{\alpha} &= \sigma_{\alpha+1}\\
\sigma_{\mu} \times \sigma_\mu &= 1 + \rho_\mu
\end{aligned}
\end{equation}
where, generalizing Eq. \eqref{eq:rho_def}, we define the fields $\rho_\mu$ by $C_{\sigma_{\mu} \sigma_{\mu}}^{\rho_\mu}=\sqrt{n\sqrt{h}}$ using the properties of the four point functions derived in Appendix \ref{sec:appendix_4_point_functions}. This also yields $C_{\rho_\mu \rho_{\mu'}}^{1}=-\frac{1}{n}$.

\section{Four point functions of spin fields}
\label{sec:appendix_4_point_functions}
The calculation of the braiding properties ultimately relies on the knowledge of the four-point functions of spin fields $\sigma_\mu$ in the $\pf{n+1}{2}{n}$ CFTs, which we present here, following \cite{Ardonne2007}. By virtue of Eq.~\eqref{eq:pfdef}, the spin fields are related to the following WZW primary fields which transform according to the fundamental representation:
\begin{equation}
\begin{aligned}
H_0 \left(w\right)&= \sigma_0 (w) e^{iq_0 \cdot \phi(w) /\sqrt{2}}\\
H_{\alpha}^{-1}\left(w\right) &= \sigma_\alpha\! \left(w\right) e^{-iq_\alpha \cdot \phi \left(w\right)/\sqrt{2}}.\\
\end{aligned}
\end{equation}
To simplify the notation, we write these as $g_\mu$, $\mu=0,1,\dots,n$ where $g_0 = H_0$ and $g_\alpha = H_{\alpha}^{-1}$. Then, $g_\mu^{-1}$ transforms according to the anti-fundamental representation. The four point functions of such WZW primaries are given by\footnote{We note that the results are obtained in Ref. \cite{Knizhnik1984} with a different convention for the anharmonic ratio. } \cite{Knizhnik1984}
\begin{align}
\label{eq:wzw_four_point_1}
C^{(p)}_1 &= \big< g_{\mu}(w_1) g_{\mu}^{-1}(w_2) g_{\mu'}^{-1} (w_3) g_{\mu'} (w_4) \big>^{\left(p\right)}\\
\nonumber
&= [w_{12}w_{34}]^{-2\Delta} x^{2\Delta} (1-x)^{-\frac{1}{n+1}} h^{\frac{p}{2}}
\mathcal{F}_{1}^{p}\left(x\right) \\
\label{eq:wzw_four_point_2}
C^{(p)}_2 &= \big< g_{\mu} (w_1) g_{\mu'}^{-1}(w_2) g_{\mu}^{-1}(w_3) g_{\mu'}(w_4)\big>^{\left(p\right)}\\
\nonumber
&= [w_{12}w_{34}]^{-2\Delta} x^{2\Delta} (1-x)^{-\frac{1}{n+1}} h^{\frac{p}{2}} [-x \mathcal{F}_1^p \left(x\right)+(1-x)\mathcal{F}_2^p \left(x\right)] \\
\label{eq:wzw_four_point_3}
C^{(p)}_1 + C^{(p)}_2 &=\big< g_{\mu} (w_1) g_{\mu}^{-1} (w_2) g_{\mu}^{-1} (w_3) g_{\mu} (w_4)\big>^{\left(p\right)}\\
\nonumber
&= [w_{12}w_{34}]^{-2\Delta} x^{2\Delta} (1-x)^{\frac{n}{n+1}}h^{\frac{p}{2}} [\mathcal{F}_{1}^{p}\left(x\right) + \mathcal{F}_{2}^{p}\left(x\right)]
\end{align}
where $p=0,1$ denotes the fusion channel and $\Delta=\frac{n(n+2)}{2(n+1)(n+3)}$ is the conformal dimension of $g$. We remind the reader of the notation $w_{ij} = w_i -w_j$ and $x=\frac{w_{12} w_{34}}{w_{13}w_{24}}$. Additionally $\sqrt{h} = \frac{1}{(n+1)} \sqrt{\frac{\Gamma(\frac{n}{n+3}) \Gamma(\frac{n+2}{n+3}) }{\Gamma(\frac{3}{n+3})\Gamma(\frac{1}{n+3})}}\frac{\Gamma(\frac{2}{n+3})}{ \Gamma (\frac{n+1}{n+3})}$ and the $\mathcal{F}_{i}^{p}$ are the following functions in terms of the hypergeometric functions $\Hypergeometric{2}{1}{a,b}{c}{x}$:
\begin{equation}
\begin{aligned}
\mathcal{F}_{1}^{0}\left(x\right)&=x^{-2\Delta}(1-x)^{\frac{1}{(n+1)(n+3)}}\Hypergeometric{2}{1}{\frac{1}{n+3},-\frac{1}{n+3}}{\frac{2}{n+3}}{x}\\
\mathcal{F}_{2}^{0}\left(x\right)&=\frac{1}{2}x^{1-2\Delta}(1-x)^{\frac{1}{(n+1)(n+3)}}
\Hypergeometric{2}{1}{1+\frac{1}{n+3},1-\frac{1}{n+3}}{1+\frac{2}{n+3}}{x}\\
\mathcal{F}_{1}^{1}\left(x\right)&=x^{\frac{1}{(n+1)(n+3)}}(1-x)^{\frac{1}{(n+1)(n+3)}}\Hypergeometric{2}{1}{\frac{n}{n+3},\frac{n+2}{n+3}}{1+\frac{n+1}{n+3}}{x}\\
\mathcal{F}_{2}^{1}\left(x\right)&=-(n+1)x^{\frac{1}{(n+1)(n+3)}}(1-x)^{\frac{1}{(n+1)(n+3)}}\Hypergeometric{2}{1}{\frac{n}{n+3},\frac{n+2}{n+3}}{\frac{n+1}{n+3}}{x}.
\label{eq:F_general}
\end{aligned}
\end{equation}
Up to a phase, the four point functions of spin fields can be found from
Eq.~\eqref{eq:wzw_four_point_1} by splitting off a correlator of vertex operators. The final result is
\begin{equation}
\begin{aligned}
\big<\sigma_{\mu} \sigma_{\mu} \sigma_{\mu \hphantom{`}} \sigma_{\mu \hphantom{`}} \big>^{\left(p\right)} &= \left(-1\right)^p [w_{12}w_{34}]^{\frac{n}{2(n+1)}-2\Delta} x^{2\Delta} (1-x)^{\frac{n}{2(n+1)}} h^{\frac{p}{2}} [\mathcal{F}_{1}^{p}\left(x\right) + \mathcal{F}_{2}^{p}\left(x\right)] \\
\big<\sigma_{\mu}\sigma_{\mu}\sigma_{\mu'} \sigma_{\mu'}  \big>^{\left(p\right)} &= \left(-1\right)^p [w_{12}w_{34}]^{\frac{n}{2(n+1)}-2\Delta} x^{2\Delta} (1-x)^{\frac{-1}{2(n+1)}} h^{\frac{p}{2}} \mathcal{F}_{1}^p \left(x\right) \\
\big<\sigma_{\mu}\sigma_{\mu'}\sigma_{\mu}\sigma_{\mu'} \big>^{\left(p\right)} &=\left(-1\right)^p [w_{12}w_{34}]^{\frac{n}{2(n+1)}-2\Delta} x^{2\Delta-\frac{1}{2}} (1-x)^{\frac{-1}{2(n+1)}} h^{\frac{p}{2}} [-x\mathcal{F}_1^p \left(x\right)+ (1-x)\mathcal{F}_2^p \left(x\right)].
\end{aligned}
\end{equation}
The precise way in which the phases were obtained requires some additional clarification. In principle, these phases can be obtained by studying the behavior of the four point functions in the limit $w_{12},w_{34}\to 0$ or $x\to 0$, using the OPEs of the spin fields. For the fusion channel $p=0$ this fixes all phases to 1. For the fusion channel $p=1$ however, the distinction between the sector $\rho$ and the fields $\rho_\mu$ introduced in Appendix~\ref{app:fusion-ope} becomes important. Namely, naively using the sector $\rho$ as the $p=1$ channel in the OPEs of the spin fields, i.e. $(\sigma_\mu \sigma_\mu)^{1} \propto C_{\sigma_\mu \sigma_\mu}^\rho \rho$, the coefficients $C_{\sigma_\mu \sigma_\mu}^{\rho}$ are found to be inconsistent. 

Using the definition $\left(\sigma_\mu \sigma_\mu\right)^{1} \propto C_{\sigma_\mu \sigma_\mu}^{\rho_\mu} \rho_\mu$ instead, the normalization $C_{\rho_\mu \rho_\mu}^{1} = 1$ determines the phase for the four point function $\langle \sigma_\mu \sigma_\mu \sigma_\mu \sigma_\mu \rangle^{(1)}$, which is $-1$. For the remaining four point functions, the OPE coefficients $C_{\rho_\mu \rho_{\mu'}}^{1}$ are not known a-priori: to fix the phases an additional limit of the master formula, given in Eq. \eqref{eq:n=3_limits1a} for $n=3$ and Eq. \eqref{eq:r_limits} in the general case, is taken. This gives a consistency condition between the expansion coefficients $A^{\left(p\right)},B^{\left(p\right)}$ which is used to fix the phases and thereby the OPE coefficients. The phases are $-1$, and the OPE coefficients read $C_{\rho_\mu \rho_{\mu'}}^{1} = -\frac{1}{n}$.

\section{Transformation properties of the $\mathcal{F}_i^{p}$}
\label{sec:appendix_F_transf}
We present the transformations of the functions $\mathcal{F}_{i}^{p}\left(x\right)$ given in Eq. \eqref{eq:F_general} under $x \to 1-x$, $x \to \frac{-x}{1-x}$ and $x \to \frac{1}{x}$.  For this, the transformation properties of the hypergeometric functions are needed, as well as contiguous relations between them. 

For the transformation $w_1\leftrightarrows w_2$, corresponding to  $x\to\frac{-x}{1-x}$, we have
\begin{equation}
\begin{aligned}
\label{eq:F_transf12_1}
\mathcal{F}_1^{p} \left(\frac{-x}{1-x}\right)&= \left(-1\right)^{\frac{(n+1)p}{n+3}-2\Delta} (1-x)^{2\Delta-\frac{1}{n+1}}\mathcal{F}_1^p \left(x\right)\\
\mathcal{F}_2^{p} \left(\frac{-x}{1-x}\right)&=  \left(-1\right)^{\frac{(n+1)p}{n+3}-2\Delta} (1-x)^{2\Delta-\frac{1}{n+1}}[-x\mathcal{F}_1^p \left(x\right) +(1-x)\mathcal{F}_2^p \left(x\right)] .
\end{aligned}
\end{equation}
For the transformation $w_2\leftrightarrows w_3$, corresponding to $x \to \frac{1}{x}$: 
\begin{equation}
\begin{aligned}
\label{eq:F_transf23_1}
\mathcal{F}_{1}^{p} \left(\frac{1}{x}\right) &= \left(-1\right)^{-\frac{(n+1)p}{n+3}-\frac{n}{(n+1)(n+3)}} x^{2\Delta-\frac{1}{n+1}} [C_{0}^{p} [x\mathcal{F}_1^0 -(1-x)\mathcal{F}_2^0] - \left(-1\right)^{\frac{2}{n+3}}C_1^p [x\mathcal{F}_1^1 - (1-x)\mathcal{F}_2^1]]\\
\mathcal{F}_{2}^{p} \left(\frac{1}{x}\right) &= \left(-1\right)^{\frac{-(n+1)p}{n+3}-\frac{n}{(n+1)(n+3)}} x^{2\Delta-\frac{1}{n+1}} [C_0^p \mathcal{F}_2^0 -\left(-1\right)^{\frac{2}{n+3}} C_1^p \mathcal{F}_2^1 ]
\end{aligned}
\end{equation}
where
\begin{equation}
\begin{aligned}
C_0^0 &= -C_{1}^{1} =\frac{1}{2\cos{\left(\frac{\pi}{n+3}\right)}}\\
C_0^{1} &=\frac{1-(C_0^0)^2}{C_1^0}=  -(n+1)\frac{\Gamma^{2}(\frac{n+1}{n+3})}{\Gamma(\frac{n}{n+3})\Gamma(\frac{n+2}{n+3})}  .
\end{aligned}
\end{equation}
Finally, for the transformation $w_1 \leftrightarrows w_3$, corresponding to $x \to 1-x$ \cite{Knizhnik1984}
\begin{equation}
\begin{aligned}
\label{eq:F_transf13_1}
\mathcal{F}_{1}^{p}(1-x)&=C_{0}^{p}\mathcal{F}_{2}^{0}\left(x\right)+C_{1}^{p}\mathcal{F}_{2}^{1}\left(x\right)\\
\mathcal{F}_{2}^{p}(1-x)&=C_{0}^{p}\mathcal{F}_{1}^{0}\left(x\right)+C_{1}^{p}\mathcal{F}_{1}^{1}\left(x\right)  .
\end{aligned}
\end{equation}
To obtain the braid behavior of the fundamental quasiholes, the following identities are also useful:
\begin{equation}
\begin{aligned}
\label{eq:braid_coeffs}
d_n&= (C_0^0)^{-1}= 2\cos\left(\frac{\pi}{n+3}\right)\\
C_0^{1}\sqrt{h} &= \frac{C_1^0}{\sqrt{h}} = -\sqrt{1-(C_0^0)^2}.
\end{aligned}
\end{equation}

\section{Rank level duality}
\label{app:rank-level-duality}
We comment on the consequences of rank-level duality, which relates the $\su{n+1}{k}$ and
$\su{k}{n+1}$ WZW theories. In particular, we consider the consequences for the correlators, and thereby the braiding behavior of
the quasiholes. In \cite{Naculich1990}, the relation between the correlators of WZW primary fields in the dual WZW theories was derived. For the present purposes, we only consider the
$\su{n+1}{2}$ and $\su{2}{n+1}$ cases. The correlators of four primary fields of the former theory are given in
Eqs.~\eqref{eq:wzw_four_point_1}-\eqref{eq:wzw_four_point_3}. The equivalent correlators for the later theory are stated here, using the convention  
$x=\frac{w_{12} w_{34}}{w_{13}w_{24}}$,  which differs from the one used in \cite{Knizhnik1984}, where these correlators were derived.
The correlators $\tilde{C}^{(p)}_a$ of the fields $g$, corresponding to the fundamental representation of the $\su{2}{n+1}$ WZW theory read
\begin{align}
\label{eq:wzwcorrtilde1}
\tilde{C}_1^{(p)} &= \bigl< g_{\mu} (w_1) g_{\mu}^{-1} (w_2) g_{\mu'}^{-1} (w_3) g_{\mu'} (w_4) \bigr>^{(p)} =
w_{12}^{-2\tilde{\Delta}}w_{34}^{-2\tilde{\Delta}} x^{2\tilde{\Delta}} \tilde{h}^{\frac{p}{2}}  \bigl[ \tilde{\mathcal{F}}^{p}_1 (x) + \tilde{\mathcal{F}}^{p}_2 (x) \bigr] \\
\label{eq:wzwcorrtilde2}
\tilde{C}_2^{(p)} &= \bigl< g_{\mu} (w_1) g_{\mu'}^{-1} (w_2) g_{\mu}^{-1} (w_3) g_{\mu'} (w_4) \bigr>^{(p)} =
w_{12}^{-2\tilde{\Delta}}w_{34}^{-2\tilde{\Delta}} x^{2\tilde{\Delta}} \tilde{h}^{\frac{p}{2}}  \bigl[ - \tilde{\mathcal{F}}^{p}_2 (x) \bigr] \\
\label{eq:wzwcorrtilde3}
\tilde{C}_1^{(p)} + \tilde{C}_2^{(p)} &= \bigl< g_{\mu} (w_1) g_{\mu}^{-1} (w_2) g_{\mu}^{-1} (w_3) g_{\mu} (w_4) \bigr>^{(p)} =
w_{12}^{-2\tilde{\Delta}}w_{34}^{-2\tilde{\Delta}} x^{2\tilde{\Delta}} \tilde{h}^{\frac{p}{2}}  \bigl[ \tilde{\mathcal{F}}^{p}_1 (x) \bigr] \ ,
\end{align}
where $\mu,\mu'$ label the weights of the fundamental (i.e., two-dimensional) representation of $\su{2}{}$.
The $p=0$ channel corresponds to the trivial intermediate channel, $(0)$, while the $p=1$ channel corresponds to $(2)$, the
adjoint (i.e., three dimensional) representation.
The tilde indicates that we deal with the $\su{2}{n+1}$ quantities instead of the $\su{n+1}{2}$ version (for the general $\su{n+1}{k}$ results, see \cite{Knizhnik1984}), that is
$\tilde{\Delta} = \frac{3}{2(n+3)}$,  $\tilde{h} = \frac{\Gamma(\frac{1}{n+3})\Gamma(\frac{3}{n+3})\Gamma(\frac{n+1}{n+3})^2}{4 \Gamma(\frac{n}{n+3})\Gamma(\frac{n+2}{n+3})\Gamma(\frac{2}{n+3})^2}$ and
\begin{align}
\tilde{\mathcal{F}}_{1}^{0} (x)
&=x^{-2\tilde{\Delta}}(1-x)^{\frac{1}{2(n+3)}}\Hypergeometric{2}{1}{\frac{1}{n+3},-\frac{1}{n+3}}{\frac{n+1}{n+3}}{x} \\
\tilde{\mathcal{F}}_{2}^{0} (x)
&=\frac{1}{n+1}x^{1-2\tilde{\Delta}}(1-x)^{\frac{1}{2(n+3)}}\Hypergeometric{2}{1}{1+\frac{1}{n+3},1-\frac{1}{n+3}}{1+\frac{n+1}{n+3}}{x} \\
\tilde{\mathcal{F}}_{1}^{1} (x)
&=x^{\frac{1}{2(n+3)}}(1-x)^{\frac{1}{2(n+3)}}\Hypergeometric{2}{1}{\frac{1}{n+3},\frac{3}{n+3}}{1+\frac{2}{n+3}}{x} \\
\tilde{\mathcal{F}}_{2}^{1} (x)
&=-2x^{\frac{1}{2(n+3)}}(1-x)^{\frac{1}{2(n+3)}}\Hypergeometric{2}{1}{\frac{1}{n+3},\frac{3}{n+3}}{\frac{2}{n+3}}{x} \ .
\end{align}
For the correlators of the $\su{n+1}{2}$ and $\su{2}{n+1}$ WZW theories, rank level duality takes the
following form \cite{Naculich1990}
\begin{equation}
\bigl( \tilde{C}_1^{(0)} + \tilde{C}_2^{(0)} \bigr) 
\bigl( C_1^{(0)} + C_2^{(0)} \bigr) 
+
\bigl( \tilde{C}_1^{(1)} + \tilde{C}_2^{(1)} \bigr) 
\bigl( C_1^{(1)} + C_2^{(1)} \bigr) 
= w_{12}^{-\frac{2n+1}{2(n+1)}}w_{34}^{-\frac{2n+1}{2(n+1)}} (1-x)^{\frac{2n+1}{2(n+1)}} \ .
\label{eq:corr-duality}
\end{equation}
Before we comment on the consequences for the braid matrices, we note that we obtained the results for the correlators
$\tilde{C}_a^{(p)}$ by taking the result from \cite{Knizhnik1984}, and transforming $x\rightarrow -\frac{x}{1-x}$, to take the different
choices for the anharmonic ratios into account. This leads to the fact that for the $\su{2}{2}$ correlators, i.e. either
$C_a^{(p)}$ or $\tilde{C}_a^{(p)}$ with $n=1$, we have that $C^{(0)}_1 = \tilde{C}^{(0)}_1$ and $C^{(0)}_2 = \tilde{C}^{(0)}_2$, but in the
$p=1$ channel they differ by a sign, $C^{(1)}_1 = -\tilde{C}^{(1)}_1$ and $C^{(1)}_2 = -\tilde{C}^{(1)}_2$.

The duality relation between the correlators Eq.~\eqref{eq:corr-duality}, implies that the braid matrices are also related.
To avoid clutter in the notation, we denote braid matrices derived from the WZW correlators by $W^{(n+1,k)}_{ij}$. 
From the explicit form of the correlator Eq.~\eqref{eq:wzwcorrtilde3}, we obtain the braid matrices $W_{23}^{(2,n+1)}$
for the exchange of $w_2\leftrightarrow w_3$,
\begin{equation}
W_{23}^{(2,n+1)} = 
\frac{(-1)^{2\tilde{\Delta}}}{d_n}
\begin{pmatrix}
1 & (-1)^{-\frac{2}{n+3}} \sqrt{d_n^2-1} \\
(-1)^{-\frac{2}{n+3}} \sqrt{d_n^2-1} & -(-1)^{-\frac{4}{n+3}}\\
\end{pmatrix}
\end{equation}
From the correlator $C^{(p)}_1+C^{(p)}_2$, Eq.~\eqref{eq:wzw_four_point_3}, we obtain the equivalent braid matrix
for $\su{n+1}{2}$
\begin{equation}
\label{eq:su2_r_braid_23}
W_{23}^{(n+1,2)} = 
\frac{(-1)^{2\Delta}}{d_n}
\begin{pmatrix}
1 & (-1)^{\frac{2}{n+3}} \sqrt{d_n^2-1} \\
(-1)^{\frac{2}{n+3}} \sqrt{d_n^2-1} & -(-1)^{\frac{4}{n+3}}\\
\end{pmatrix}
\end{equation}
These matrices satisfy
\begin{equation}
\label{eq:braid-duality}
W_{23}^{(2,n+1)}\cdot  W_{23}^{(n+1,2)} = -(-1)^{-\frac{1}{2(n+1)}} \mathbf{1} \ ,
\end{equation}
as expected from the duality relation Eq.~\eqref{eq:corr-duality}, see \cite{Naculich1990}.

The matrices $W_{23}^{(n+1,2)}$ differ from the ones obtained using the parafermion correlators in Sec.~\ref{sec:r_braiding} by a sign of
the off-diagonal elements, see Eq.~\eqref{eq:r_braid23}. From an anyon-model point of view \cite{kitaev2006}, this sign is a gauge convention.
However, the (sign) `choices' made in
Sec.~\ref{sec:r_braiding} came from various consistency conditions. These choices are consistent with the choices made in \cite{Ardonne2007} in
the case $\su{3}{2}$, so indeed, the braid matrices are the same
(after taking the different choices for the anharmonic ratio into account).
In addition, these choices also coincide with natural phase choices when one calculates the
$F$- and $R$-matrices of the anyon-models using quantum groups, as explained in \cite{Ardonne_2010}.
The braid matrices $U_{23}^{(2,n+1)}$ calculated in \cite{Ardonne2007} are the same as the ones obtained from the WZW correlator
Eq.~\eqref{eq:wzw_four_point_3} (again after taking the different anharmonic ratio into account), so
$W_{23}^{(2,n+1)} = U_{23}^{(2,n+1)}$. They also correspond to the braid matrices obtained using quantum groups.

Thus, because of the difference between $W_{23}^{(n+1,2)}$ and $U_{23}^{(n+1,2)}$, it is interesting to investigate if the
braid matrices $U^{(n+1,2)}_{23}$ and $U^{(2,n+1)}_{23}=W^{(2,n+1)}_{23}$
as given in Eqs.~\eqref{eq:r_braid23} and \eqref{eq:su2_r_braid_23} are also related in some way.
Such a relation indeed exists, if one swaps the rows and
columns of $U^{(2,n+1)}_{23}$. This swap is natural, because the two fusion channels of two fundamental representations $\omega_1$ are
$(2,0)$ and $(0,1)$ for $\su{3}{2}$; $(2,0,0)$ and $(0,1,0)$ for $\su{4}{2}$;
$(2,0,0,0)$ and $(0,1,0,0)$ for $\su{5}{2}$, etc. Form this point of view, the natural ordering for
$\su{2}{2}$ would be $(2)$ and $(0)$, which is the opposite ordering in comparison to one used for
$U^{(2,n+1)}_{23}$ of Eq.~\eqref{eq:su2_r_braid_23}. We denote the version of $U^{(2,n+1)}_{23}$ with swapped rows and columns by
$\overline{U}^{(2,n+1)}_{23}$. One then easily obtains the relation
\begin{equation}
\label{eq:braid-duality2_23}
(-1)^{\frac{1}{(n+1)(n+3)}} U_{23}^{(n+1,2)} = (-1)^{\frac{1}{2(n+3)}} \overline{U}_{23}^{(2,n+1)} \ .
\end{equation}
One finds that relation between the braid matrices for exchanging $w_1\leftrightarrow w_2$ is the same, and the one for
$w_1\leftrightarrow w_3$ easily follows,
\begin{align}
\label{eq:braid-duality2_12-13}
(-1)^{\frac{1}{(n+1)(n+3)}} U_{12}^{(n+1,2)} &= (-1)^{\frac{1}{2(n+3)}} \overline{U}_{12}^{(2,n+1)} \\
(-1)^{\frac{3}{(n+1)(n+3)}} U_{13}^{(n+1,2)} &= (-1)^{\frac{3}{2(n+3)}} \overline{U}_{13}^{(2,n+1)}
\ .
\end{align}

\end{appendix}

\end{document}